%% file: ipp.tex
\documentstyle[12pt]{article}

\newcommand{\beq}{\begin{equation}}
\newcommand{\eeq}{\end{equation}}

\newcommand{\PU}{\Phi_0}
\def\nn{\nonumber\\}
\def\P{\Phi_0}
\def\A{{\bf A}}
\def\V{{\bf V}}
\def\G{{\bf G}}
\def\H{{\bf H}}
\def\C{{\bf C}}

\begin{document}
\def\lot{\ln{\Lambda\over T}}
\def\ltt{\left(\lot\right)^2}
\def\lag{\langle}
\def\rag{\rangle}
\def\noct{{CT\!\!\!\!\!\!\!\!\!\rule[0.5 ex]{1em}{0.5pt}}\hskip.5em}

\title{Partial Path Integration\\
 of Quantum Fields:\\
Two-Loop Analysis of the SU(2) \\
Gauge-Higgs Model\\
at Finite Temperature}

\author{{A. Jakov\'ac$^{1}$ and A. Patk\'os$^{2}$}\\
{Department of Atomic Physics}\\
{E\"otv\"os University, Budapest, Hungary}\\
}
\vfill
\footnotetext[1]{{\em e-mail address: jako@hercules.elte.hu, 
from September 1996: DESY, Theory Group}}
\footnotetext[2]{{\em e-mail address: patkos@ludens.elte.hu}}
\maketitle
\begin{abstract}
\noindent
High temperature reduction of the SU(2) Higgs model is realised by partially
integrating its partition function. Various approximate forms of the effective
theory resulting from the integration over nonstatic fields and the static
electric potential are analysed. Also non-polynomial and non-local
terms are allowed. Consistency of the perturbative solution is ensured
by new types of induced counterterms. Perturbative phase transition
characteristics are presented in the Higgs mass range 30-120 GeV, and 
compared to results of other perturbative approaches.
\end{abstract}
\newpage
\section{Introduction}
Physical systems develop at finite temperature multiple 
mass-scales. Decoupling theorems \cite{App75} ensure that effective 
representations 
can be found for the lighter degrees of freedom at scales asymptotically 
separated from the heavier scales. The effective couplings can be related 
to the parameters of the original system by matching values of the
same quantities calculated in the effective model and the original model
simultanously \cite{Wei80,Geo93,Bra95}. 

In weakly coupled gauge theories (the class where the electroweak theory 
belongs to)  matching can be performed perturbatively. 
The simplest is to match the renormalised Green's functions
computed in the two theories separately. This tacitly assumes the 
renormalisability of the effective model. Though at large distances
local operators of lower scaling dimensions dominate indeed, it is important
to assess quantitatively the rate of suppression of higher dimensional
operators. 

In finite temperature theories integration over non-static fields
yields for static $n$-point functions with typical momenta ${\cal O}(p)$,
expressions analytic in $p^2/M_{heavy}^2$, therefore the occuring
non-localities can always be expanded in a kind of gradient expansion. 
We shall demonstrate, that in Higgs systems to
${\cal O}(g^4,\lambda g^2,\lambda^2$) accuracy, the non-local effects 
in the effective Higgs-potential arising from non-static fields 
can be represented {\it exactly} by correcting the
coefficient of the term quadratic in the Higgs-field.

Similar conclusions are arrived at when the contribution of non-static modes 
to higher dimensional (non-derivative) operators is investigated.

The situation changes when a heavy {\it static} field is 
integrated out. The contribution to the polarisation functions from this 
degree of freedom is not analytic neither in  $p^2/M_{heavy}^2$ nor 
in $\Phi^2/M_{heavy}^2$. For the momentum dependence no gradient expansion  
can be proven to exist.
Though the contribution to the effective Higgs-potential can be represented
approximately by an appropriately corrected local gauge-Higgs
theory, it does not automatically correspond to a systematic weak coupling 
expansion. Assuming that the important range of variation of the Higgs 
field is $\sim gT$, one proceeds with the expansion, and the validity of this 
assumption is checked {\it a posteriori}. In case of strong first order 
transitions this assumption is certainly incorrect, but seems to be justified 
for the Standard Model with Higgs masses around 80GeV.

Non-analytic dependence on field variables is reflected in the 
appearence of non-polynomial local terms in the reduced effective theory. 
Consistent treatment of such pieces is an interesting result of our paper 
(see the discussion of the U(1) case in \cite{Jak96b}).

The method of integration over specific (heavy static and non-static) 
classes of fields in the path integral (Partial Path Integration, PPI), 
instead of matching,
provides us with a direct method of inducing all non-local and non-polynomial
corrections automatically. Though the calculations are much more painful, 
than those involved in matching, the value of PPI in our eyes is just the 
possibility of assessing the range of applicability of the technically 
simpler method. 

The explicit expression of the non-local and/or non-polynomial 
parts of the action and the impact of these
terms on the effective finite temperature Higgs-potential will be 
investigated in the present paper for the SU(2) gauge-Higgs model. 
This model is of central physical importance in investigating the nature of 
the  cosmological electroweak phase transition \cite{Sin95}. Earlier, PPI has 
been applied to the O(N) model by one of us \cite{Jak96}. 

A similar, but even more ambitious program is being pursued by Mack and
collaborators \cite{Ker95}. They attempt at performing PPI with the goal of
deriving in a unique procedure the perfect lattice action of the coarse 
grained effective model for the light degrees of freedom.

Our calculation stays within the continuum perturbative framework. 
PPI will be performed with two-loop accuracy for two reasons:

i) The realisation of PPI on Gaussian level does not
produce non-local terms, though non-polynomial local terms already do appear. 
The consistent cut-off independence of the perturbation theory
with non-polynomial terms can be tested first in 
calculations which involve also 2-loop diagrams.

ii) It has been demonstrated that the quantitative non-perturbative 
characterisation of the effective SU(2) gauge-Higgs
model is sensitive to two-loop corrections of the cut-off (lattice
spacing) dependence of its effective bare couplings \cite{Kaj95,Kaj96}. 
We would like to investigate possible variations of the relation of the
effective couplings to the parameters of the original theory when the applied
regularisation and renormalisation schemes change.

The integration over the non-static and of the heavy static fields will not
be separated, but is performed in parallel with help of the thermal 
counterterm technique already applied in the PPI of the U(1) Higgs model
\cite{Jak96b}.

The paper is organised in the following way. In section 2 the basic 
definitions are given and the operational steps involved in
the 2-loop PPI are listed. In Section 3 we are going to discuss the
contributions from fully non-static fluctuations. In Section 4 the 
contribution of the diagrams involving also heavy static propagators is
discussed. In both sections particular attention is payed to the analysis of 
the effective non-local interactions. In Section 5 the effective Higgs 
potential is calculated from the 2-loop level effective 
non-local and non-polynomial 3-d (NNA) action with 2-loop accuracy.
Also the local polynomial (LPA) and local non-polynomial (LNA) approximations
are worked out. The quantitative perturbative characterisation of the phase
transition and its comparison with other approaches will be presented 
in Section 6. Our 
conclusions are summarised in Section 7. In order to make the paper better
readable, the details of the computations discussed in Sections 3 to 5 are 
relegated to various Appendices. 

\section{Steps of perturbative PPI}

The model under consideration is the SU(2) gauge-Higgs model with one complex
Higgs doublet:
\begin{eqnarray}
&
{\cal L}={1\over 4} F_{mn}^aF_{mn}^a
+ {1\over 2} (\nabla_m\Phi)^{\dagger}(\nabla_m\Phi)
+ {1\over 2} m^2 \Phi^\dagger\Phi + {\lambda\over24}(\Phi^\dagger\Phi)^2
+ {\cal L}_{CT},\nonumber\\
&
F_{mn}^a  =\partial_m A_n^a -\partial_n A_m^a + g \epsilon_{abc}
A_m^bA_n^c, \nonumber\\
&
\nabla_m  =\partial_m - igA_m^a\tau_a.
\end{eqnarray}
The integrations over the non-static Matsubara modes and the static electric
component of the vector potential will be performed with two-loop accuracy.
In the case of the static electric potential resummed perturbation theory 
is applied, simply realised
by adding the mass term $(Tm_D^2/2)\int d^3x\left(A_0^a({\bf x})\right)^2$
to the free action and compensating for it in the interaction piece. 
The resummation discriminates between the static and non-static $A_0$ 
components, therefore in the Lagrangean density the replacement
\begin{equation}
A_0({\bf x},\tau )\rightarrow A_0({\bf x})+a_0({\bf x},\tau )
\end{equation}
should be done. (With lower case letters we always refer to non-static
fields.)

In the first step we are going to calculate the local (potential) part of the
reduced Lagrangean. For this a constant background is introduced into the
Higgs-multiplet:
\begin{eqnarray}
&
\Phi ({\bf x},\tau )=\Phi_0+\phi ({\bf x},\tau ),\nonumber\\
&
\phi ({\bf x},\tau )=\left(\matrix{\xi_4+i\xi_3\cr
i\xi_1-\xi_2\cr}\right),
\end{eqnarray}
where $\Phi_0$ was chosen to be the constant counterpart of $\xi_4$.
The dependence on the other static components is reconstructed by requiring
the O(4) symmetry of the potential energy.

In the second step the kinetic piece of the reduced Lagrangean is
investigated. In order to fix the coefficient of the conventional kinetic
piece to 1/2, one has to rescale the fields. The wave function
renormalisation constants should be calculated with
${\cal O}(g)$ accuracy. The derivative part of the scalar
action can be extended upon the requirement of spatial gauge invariance
to include the magnetic vector-scalar interaction into itself.
     
Higher derivative contributions to the effective action can be summarized
under the common name {\it non-local} interactions. They appear first in the
two-loop calculation of the polarisation functions of the different fields.

The exponent of the functional integration has the following explicit 
expression:
\begin{eqnarray}
&
 {\cal L}^{(cl)}+{\cal L}^{(2)}= & {1\over 2} m^2\Phi_0^2+{\lambda\over24}\PU^4 \nonumber\\
&&+{1\over 2} a_m^a(-K)\left[\left(K^2+{g^2\over4}\PU^2\right)\delta_{mn}-
\left(1-{1\over\alpha}\right)K_mK_n\right]a_n^a(K)\nonumber\\
&&+{1\over 2} A_0^a(-k)\left[k^2+{g^2\over4}\PU^2+m_D^2\right]A_0^a(k)
\nonumber\\
&&+{1\over 2}\xi_H(-K)\left[K^2+m^2+{\lambda\over2}\PU^2\right]\xi_H(K)
\nonumber\\
&&
+{1\over 2}\xi_a(-K)\left[K^2+m^2+{\lambda\over6}\PU^2\right]\xi_a(K)
+c_a^\dagger(K)K^2c_a(K).
\end{eqnarray}
where $c$ denotes the ghost fields, and the notation $\xi_4\equiv\xi_H$ has 
been introduced.  Landau gauge fixing and the
Faddeev-Popov ghost terms are included. 
The above expression implies the following propagators for the different fields
to be integrated out:
\begin{eqnarray}
& \langle a_m^a(P)a_n^b(Q) \rangle = {\displaystyle\delta_{ab}\over
\displaystyle P^2+m_a^2}
\left(\delta_{mn}-{\displaystyle P_mP_n\over\displaystyle P^2}\right)
\hat\delta(P+Q), & m_a^2={g^2\over 4}\PU^2,\nonumber\\
& \langle A_0^a(P)A_0^b(Q) \rangle = {\displaystyle\delta_{ab}\over
\displaystyle P^2+m_{A0}^2}\hat\delta(P+Q), 
& m_{A0}^2=m_D^2+{g^2\over 4}\PU^2,\nonumber\\
& \langle \xi_H(P)\xi_H(Q) \rangle ={\displaystyle1\over
\displaystyle P^2+m_H^2}\hat\delta(P+Q) ,
 & m_H^2=m^2+{\lambda\over2}\PU^2,\nonumber\\
& \langle \xi_a(P)\xi_b(Q) \rangle ={\displaystyle\delta_{ab}\over
\displaystyle P^2+m_G^2}
\hat\delta(P+Q), & m_G^2=m^2+{\lambda\over6}\PU^2,\nonumber\\
& \langle c_a^\dagger(P)c_b(Q) \rangle = -{\displaystyle\delta_{ab}\over
\displaystyle P^2}\hat\delta(P-Q).&
\end{eqnarray}
($\hat\delta$ denotes the finite temperature generalisation of Dirac's
delta-function.)

The interaction part of the Lagrangean density consists of two
parts. In the first set all non-quadratic vertices are collected:
\begin{eqnarray}
&{\cal L_I}= & igS_{abc}^{mnl}(P,Q,K)a_m^a(P)a_n^b(Q)a_l^c(K)\nonumber\\
&&+
3igS_{abc}^{mn0}(P,Q,k)
a_m^a(P)a_n^b(Q)A_0^c(k)\nonumber\\
&&+{g^2\over4}\left[(a_m^aa_m^a)^2-(a_m^aa_n^a)^2+2(A_0^aA_0^aa_i^ba_i^b
-A_0^aA_0^ba_i^aa_i^b)\right]\nonumber\\
&& +g^2(A_0^aa_0^aa_i^ba_i^b-A_0^aa_0^ba_i^aa_i^b)
\nonumber\\
&& -i{g\over2}a_m^a(P)\left((K-Q)_m\xi_a(Q)\xi_H(K)
+\epsilon_{abc}K_m\xi_b(Q)\xi_c(K)\right)\nonumber\\
&& -i{g\over2}A_0^a(p)\left((K-Q)_0\xi_a(Q)\xi_H(K)
+\epsilon_{abc}K_0\xi_b(Q)\xi_c(K)\right)\nonumber\\
&& +{g^2\over8}(a_m^aa_m^a+2A_0^aa_0^a+A_0^aA_0^a)(\xi_H^2+\xi_a^2)
+{g^2\over4}\PU\xi_H(a_m^aa_m^a+2A_0^aa_0^a)\nonumber\\
&& +{\lambda\over24}(\xi_H^2+\xi_a^2)^2
+{\lambda\over6}\PU\xi_H(\xi_H^2+\xi_a^2)
+ig\epsilon_{abc}P_mc_a^\dagger(P)a_m^b(Q)c_c(K)\nonumber\\
&&
+ig\epsilon_{abc}P_0c_a^\dagger(P)A_0^b(q)c_c(K),
\label{eq:pert}
\end{eqnarray}
where the symmetrised trilinear coupling is
\begin{equation}
S_{abc}^{mnl}(P,Q,K)={1\over6}\epsilon_{abc}\left[(P-Q)_l\delta_{mn}+
(Q-K)_m\delta_{nl}+(K-P)_n\delta_{lm}\right].
\end{equation}
The second piece is quadratic and corresponds to the $T=0$ and the thermal
counterterms:
\begin{eqnarray}
&{\cal L}_{CT}= & {1\over 2}(Z_\Phi\delta m^2+(Z_\Phi-1)m^2)\PU^2
+(Z_\Phi^2Z_\lambda-1){\lambda\over24}\PU^4\nonumber\\
&&
+{1\over 2}(Z_A-1)a_m^a\left(K^2\delta_{mn}-\left(1-{1\over\alpha}\right)K_mK_n
\right)a_n^a\nonumber\\
&& +{g^2\over8}(Z_AZ_\Phi Z_g^2-1)\PU^2a_m^2
+{1\over 2} (Z_A-1)A_0^a k^2 A_0^a \nonumber\\
&&
+{g^2\over8}(Z_AZ_\Phi Z_g^2-1)\PU^2A_0^2
-{1\over 2}m_D^2 A_0^2 \nonumber\\
&& +{1\over 2}\xi_H\left((Z_\Phi-1)(K^2+m^2)+Z_\Phi\delta m^2
+(Z_\Phi^2Z_\lambda-1){\lambda\over2}\PU^2\right)\xi_H\nonumber\\
&& +{1\over 2}\xi_a\left((Z_\Phi-1)(K^2+m^2)+Z_\Phi\delta m^2
+(Z_\Phi^2Z_\lambda-1){\lambda\over6}\PU^2\right)\xi_a\nonumber\\
&&
+(Z_c-1)c_a^\dagger K^2c_a.
\label{Lct}
\end{eqnarray}
The multiplicative and additive renormalisation constants are defined from
the relations between the renormalised and the bare couplings, as listed next:
\begin{eqnarray}
& g_B=Z_g g, & \lambda_B=Z_\lambda\lambda ,\nonumber\\
  & A_B=Z_A^{1/2} A, & \Phi_B=Z_\Phi^{1/2} \Phi ,\nonumber\\
& c_B=Z_c c, & m_B^2=m^2+\delta m^2.
\end{eqnarray}
The quadratic approximation to the counterterms is sufficient 
for our calculation, since they contribute only at one-loop.

The new couplings, emerging after the 4-d (T=0) renormalisation conditions
are imposed, are finite from the point of view of the 4-d
ultraviolet behavior, but should be considered to be the bare couplings
of the effective field theory, since they explicitly might depend on
the 3-d cut-off.

\section{Non-static fluctuations}     

\subsection{The potential (local) term of the effective action}

The 1-loop contribution to the potential energy of a homogenous 
$\PU$-background is expressed through the standard sum-integral
(with the $n=0$ mode excluded from the sum)
\begin{equation}
I_4(m)=\int_K^\prime\ln (K^2+m^2).
\end{equation}
The meaning of the primed integration symbol and the characteristic features 
of this integral are detailed in Appendix A.
Also the counterterms evaluated with the background $\Phi_0$ belong to the 
1-loop potential. The complete 1-loop expression of the regularised bare
potential is given by
\begin{eqnarray}
V_{\noct}^{(1)} &=& {1\over 2}\PU^2
	\biggl[\left({9g^2\over 4} +\lambda\right){\Lambda^2\over 8\pi^2}
   +\left({3 g^2\over 16} + {\lambda\over 12} \right) T^2\nonumber\\
&& 
-\left( {3 g^2\over 2} + \lambda\right){\Lambda T\over 2\pi^2}
 +\left(1 + 2 I_{20}\lambda - {\lambda\over 8\pi^2}\ln{\Lambda\over T}\right) 
	m^2\biggr]\nonumber\\
&& 
+{1\over 24}\PU^4\biggl[\lambda +
   \left({27g^4\over 4}+4\lambda^2\right)
      \left(I_{20}-{1\over 16\pi^2}\ln{\Lambda\over T}\right)\biggr]
\label{eq:1leffact}
\end{eqnarray}
(for the meaning of the notation $I_{20}$, and others below, see Appendix A).
Its temperature independent cut-off dependences are cancelled by 
appropriately choosing the coefficients of the renormalisation constants
in the "classical" counterterm expression:
\begin{equation}
V_{CT}^{(1)}={1\over 2}(\delta m_1^2+Z_{\Phi1}m^2)\PU^2
+(2Z_{\Phi1}+Z_{\lambda1}){\lambda\over24}\PU^4.
\end{equation}

The list of two-loop diagrams and the algebraic expressions for each of them
is the same as in case of the effective potential calculation. They appear in
several papers \cite{Arn93,Fod94} in terms of two standard sum-integrals, 
in addition to the $I_4$ function defined above.
These functions in the present case are analytic in the propagator masses 
$m_i^2$ because the zero-frequency modes are excluded from the summation:
\begin{eqnarray}
H_4(m_1,m_2.m_3) &=
\int_{P1}^\prime\int_{P2}^\prime\int_{P3}^\prime\delta (P_1+P_2+P_3)\nonumber\\
&{\displaystyle 1\over 
\displaystyle (P_1^2+m_1^2)(P_2^2+m_2^2)(P_3^2+m_3^2)},\nonumber\\
L_4(m_1,m_2) &=\int^\prime_{P1}\int^\prime_{P2}{\displaystyle (P_1P_2)^2\over 
\displaystyle P_1^2(P_1^2+m_1^2)P_2^2(P_2^2+m_2^2)}.
\end{eqnarray}
The latter expression cancels algebraically from the sum of the 2-loop 
contributions. The main properties of the functions $H_4$ and $L_4$ 
appear in Appendix A. The sum of the two-loop contributions 
without the counterterms, 
displaying typical three-dimensional linear and logarithmic cut-off 
dependences leads to the following regularised expression (for the values of 
the constants $K_{..}$ and $I_{..}$, see Appendix A):
\begin{eqnarray}
V^{(2)}_{\noct} &=& \PU^4\,\Biggl\{\,
     g^6 \left({21 I_{20}^2\over 16} + {15 I_3\over 64}
   + {87 K_{10}\over 128}\right)
   + g^4 \lambda\left({33 I_{20}^2\over 32} - {K_{10}\over64}\right) \nonumber\\
&& + g^2 \lambda^2 \left({3 I_3\over 32} + {K_{10}\over8}\right)
   + \lambda^3 \left({5 I_{20}^2\over 18} +{I_3\over24}
   - {7 K_{10}\over 108}\right)\nonumber\\
&& + \lot \Biggl[g^2 \lambda^2 {K_{1log}\over 8}
   + g^6 \left({87 K_{1log}\over 128} - {21 I_{20}\over128 \pi^2}\right)
\nonumber\\
&& + g^4 \lambda \left(-{K_{1log}\over64}-{33I_{20}\over256\pi^2}\right)
   + \lambda^3 \left(-{7 K_{1log}\over 108}-{5I_{20}\over144\pi^2}\right)
     \Biggr]\nonumber\\
&& + \ltt\Biggl[-{177 g^6\over16384 \pi^4} + {9 g^4 \lambda\over2048 \pi^4}
   - {3 g^2 \lambda^2\over1024 \pi^4}
   + {\lambda^3\over384 \pi^4}\Biggr]\Biggr\}\nonumber\\
&+&\!\!\PU^2\,\Biggl\{\,
 \Lambda^2\Biggl[\lambda^2 \left(-{K_{02}\over6} + {I_{20}\over8 \pi^2}\right)
   + g^2 \lambda \left({3 K_{02}\over 4} + {9 I_{20}\over32 \pi^2}\right)
\nonumber\\
&& + g^4 \left({81 K_{02}\over 32}
   + {15 I_{20}\over16 \pi^2}\right)\Biggr]
   + \Lambda^2\lot\Biggl[-{15 g^4\over256 \pi^4} -
	{9 g^2 \lambda\over512 \pi^4}
   - {\lambda^2\over128 \pi^4}\Biggr]\nonumber\\
&& + \Lambda T \Biggl[g^4 \left({81 K_{01}\over 32}-{15I_{20}\over4\pi^2}\right)
   + g^2 \lambda \left({3 K_{01}\over 4} - {9 I_{20}\over8 \pi^2}\right)
\nonumber\\
&& + \lambda^2 \left(-{K_{01}\over6} - {I_{20}\over2 \pi^2}\right)\Biggr]
   + \Lambda T\lot \Biggl[{15 g^4\over64 \pi^4} + {9 g^2 \lambda\over128 \pi^4}
   + {\lambda^2\over32 \pi^4}\Biggr]\nonumber\\
&& + T^2\Biggl[g^4\left({5 I_{20}\over 8} + {81 K_{00}\over 32}\right)
   + \lambda g^2\left({3 I_{20}\over 16} +{3 K_{00}\over 4}\right)\nonumber\\
&& + \lambda^2 \left({I_{20}\over12} - {K_{00}\over6}\right)\Biggr]
   + T^2\lot\Biggl[{365 g^4\over1024 \pi^2} + {27 g^2 \lambda\over256 \pi^2}
   - {\lambda^2\over32 \pi^2}\Biggr]\Biggr\}
\end{eqnarray}
From this expression beyond the constants, also terms proportional to $m^2\PU^2$
are omitted. For the required accuracy $m^2$ can be related on the tree level 
to the $T=0$ Higgs-mass. Since this mass is proportional to $\lambda$, the 
terms proportional to $m^2$ are actually ${\cal O}(g^4\lambda ,
g^2\lambda^2 ,\lambda^3)$.

One also has  to compute the 1-loop non-static counterterm contributions, 
whose sum is simply:
\begin{eqnarray}
V_{CT} &= & {9\over2} m_a^2(Z_{\Phi1}+2Z_{g1})I_4(m_a)\nonumber\\
&&
+ {1\over2}\left({\lambda\over2}\PU^2(Z_{\lambda1}
+Z_{\Phi1})+\delta m_1^2\right)I_4(m_a)\nonumber\\
&&
+{3\over2}\left({\lambda\over6}\PU^2(Z_{\lambda1}
+Z_{\Phi1})+\delta m_1^2\right)I_4(m_a).
\end{eqnarray}

The renormalised potential term of the effective theory can be determined 
once the wave function renormalisation constants are known, by choosing 
expressions for $\delta m^2$ and $\delta Z_{\lambda}$ to fulfill some
temperature independent renormalisation conditions  for the potential energy.
So, we need the wave function rescalings to 1-loop accuracy first.

\subsection{Kinetic terms of the effective theory}

The effective kinetic terms are extracted from the gradient expansion of
appropriately chosen two-point functions. More closely,
they are determined by the coefficients of the linear term in the expansion 
of the 2-point functions into power series with respect to $p^2$.
Two kinds of diagrams appear at one-loop level. The tadpole-type is momentum 
independent. The bubble diagrams are listed in Appendix B accompanied by the
corresponding analytic expressions, expanded to $p^2$ terms.

One adds to the corresponding analytic expressions the "classical value" of 
the counterterms. The renormalisation
 constants $Z$ are fixed by requiring unit residue for the propagators:
 \begin{eqnarray}
Z_A &=& 1+{25g^2\over48\pi^2}D_0 - {281g^2\over720\pi^2},\nonumber\\
Z_\Phi &=& 1+{9g^2\over32\pi^2}D_0-{g^2\over4\pi^2}
\label{z1}
\end{eqnarray}
($D_0=\ln (\Lambda /T)-\ln 2\pi +\gamma_E$).
The gauge coupling renormalisation is found by requiring that the gauge 
mass term proportional to $\PU^2$, generated radiatively, vanish 
(in other words the coupling in front of $(A_i^a\PU )^2$ stays at $g^2$):
\begin{equation}
Z_g =1-{43g^2\over96\pi^2}D_0+{701g^2\over1440\pi^2}+{\lambda\over48\pi^2}.
\label{zcoupling}
\end{equation}

In terms of the renormalised fields the kinetic term of the effective 
Lagrangian can be written down by using its
invariance under spatial gauge transformations in the usual form:
\begin{equation}
L_{kin}={1\over 4}F_{ij}^aF_{ij}^a+{1\over 2}(\nabla_i\Phi )^\dagger
\nabla_i\Phi,\quad\quad i,j=1,2,3.
\end{equation}

\subsection{Renormalisation}

Now, we can write after applying (\ref{z1}) to the $\PU$-field 
that form of the regularised potential energy, where
the remaining cut-off dependence of 4-dimensional character is due 
exclusively to the mass and scalar self-coupling renormalisation:
\begin{eqnarray}
V^{(2)} &=&
\PU^4\,\Biggl\{\, \lambda^3\left(-{19 I_{20}^2\over18} + {I_3\over24}
  - {7 K_{10}\over 108}\right)
  + g^4 \lambda \left(-{39 I_{20}^2\over32} - {K_{10}\over64}
  + {3 I_{20}\over128\pi^2}\right)\nonumber\\
&&+ g^2 \lambda^2 \left({3 I_{20}^2\over2} + {3 I_3\over32}
  + {K_{10}\over8} - {I_{20}\over96\pi^2}\right)\nonumber\\
&&+ g^6 \left({219 I_{20}^2\over32} + {15 I_3\over64} + {87 K_{10}\over 128}
  + {157 I_{20}\over2560 \pi^2}\right)\nonumber\\
&&+ \lot \Biggl[ g^6\left({87 K_{1log}\over128} - {157\over 40960 \pi^4}
  - {219 I_{20} \over256 \pi^2}\right) \nonumber\\
&&+ g^2 \lambda^2\left({K_{1log}\over 8}+{1\over1536\pi^4}
  - {3 I_{20}\over16\pi^2}\right)+ \lambda^3 \left(-{7 K_{1log}\over108}
  + {19 I_{20} \over144\pi^2}\right)\nonumber\\
&&+ g^4 \lambda \left(-{K_{1log}\over 64} - {3\over2048 \pi^4}
  + {39 I_{20}\over256 \pi^2}\right)\Biggr]\nonumber\\
&&+ \ltt \Biggl[{177 g^6\over16384 \pi^4}
  - {9 g^4 \lambda\over2048 \pi^4} + {3 g^2 \lambda^2\over1024 \pi^4}
  - {\lambda^3\over384 \pi^4}\Biggr]\Biggr\} \nonumber\\
&+&\!\!\PU^2\,\Biggl\{\,
 \Lambda T \Biggl[
    g^4 \left({81 K_{01}\over 32} - {157\over2560 \pi^4}
 - {243 I_{20}\over32\pi^2}\right)
  + \lambda^2 \left(-{K_{01}\over 6} + {I_{20}\over2 \pi^2}\right)
\nonumber\\
&&+ g^2 \lambda \left({3 K_{01}\over 4} - {1\over64\pi^4}
  - {9 I_{20}\over4\pi^2}\right)\Biggr]
  + \Lambda T\lot \Biggl[{243 g^4\over512\pi^4} +{9 g^2 \lambda\over64\pi^4}
  - {\lambda^2\over32\pi^4} \Biggr]\nonumber\\
&&+ \Lambda^2 \Biggl[
    \lambda^2 \left(-{K_{02}\over 6} - {I_{20}\over4 \pi^2}\right)
  + g^2 \lambda \left({3 K_{02}\over 4} + {1\over256 \pi^4}
  + {9 I_{20}\over32 \pi^2} \right)\nonumber\\
&&+ g^4 \left({81 K_{02}\over 32} +{157\over 10240 \pi^4}
  + {243 I_{20}\over128 \pi^2}\right)\Biggr]\nonumber\\
&&+ \Lambda^2\lot\Biggl[ - {243 g^4\over 2048 \pi^4}
  - {9 g^2 \lambda\over512 \pi^4} + {\lambda^2\over 64 \pi^4}\Biggr]\nonumber\\
&&+ T^2 \Biggl[
    \lambda^2 \left(-{I_{20}\over 12} - {K_{00}\over 6}\right)
  + g^2 \lambda \left({3 I_{20}\over 8} + {3 K_{00}\over 4}
  + {1\over 384 \pi^2}\right)\nonumber\\
&&+ g^4 \left({81 I_{20}\over 64} + {81 K_{00}\over 32}
  + {157\over15360 \pi^2}\right)\Biggr]\nonumber\\
&&+ T^2\lot\Biggl[
    {81 g^4\over256 \pi^2} + {3 g^2 \lambda\over 32 \pi^2}
   - {\lambda^2\over48 \pi^2}\Biggr]\Biggr\}.
\end{eqnarray}

The final step is to  fix the
parameters of the renormalised potential energy by enforcing certain
renormalisation conditions. We are going to use the 
simplest conditions, fixing the second and fourth derivatives of the
temperature independent part od the potential energy at the origin:
\begin{eqnarray}
&
{d^2V(T-independent)\over d\PU^2}=-m^2,\nonumber\\
&
{d^4V(T-independent)\over d\PU^4}=\lambda.
\label{rencond}
\end{eqnarray}
One pays for this the price of having more  complicated relations to the
$T=0$ physical observables. The connection of the Higgs and of the vector 
masses as well as the vacuum expectation value of the Higgs field to
the couplings renormalised through the above conditions are given in Appendix 
C.

The renormalised potential term of the reduced model is finally given by
\begin{eqnarray}
V & = &{1\over 2} m(T)^2 \Phi^\dagger\Phi + 
{\lambda \over24}(\Phi^\dagger\Phi)^2,\nonumber\\
m(T)^2 & = & m^2 + T^2\biggl[{3\over16}g^2+{1\over12}\lambda
+g^4\left({81\over32}I_{20}+{81\over16}K_{00}
+{157\over7680\pi^2}\right)\nonumber\\
&&
+g^2\lambda\left({3\over4}I_{20}
+{3\over2}K_{00}+{1\over192\pi^2}\right)
-\lambda^2\left({1\over6}I_{20}+{1\over3}K_{00}\right)\biggr]\nonumber\\
&& +\Lambda T \Biggl[
    g^4 \left({81 K_{01}\over 32} - {157\over2560 \pi^4}
  - {243 I_{20}\over32\pi^2}\right)
  + \lambda^2 \left(-{K_{01}\over 6} + {I_{20}\over2 \pi^2}\right)
\nonumber\\
&&+ g^2 \lambda \left({3 K_{01}\over 4} - {1\over64\pi^4}
  - {9 I_{20}\over4\pi^2}\right)\Biggr]
  + \Lambda T\ln{\Lambda\over T} 
\Biggl[{243 g^4\over512\pi^4} +{9 g^2 \lambda\over64\pi^4}
  - {\lambda^2\over32\pi^4} \Biggr]\nonumber\\
&&+ T^2\ln{\Lambda\over T}\Biggl[
    {81 g^4\over128\pi^2} + {3 g^2 \lambda\over 16 \pi^2}
  - {\lambda^2\over24 \pi^2}\Biggr].
\label{bare3dmass}
\end{eqnarray}
 The last term of (\ref{bare3dmass}) can be split
into the sum of a finite and an infinite term by introducing a 3d scale $\mu_3$
into it. It is very remarkable, that the 3d scale dependence at this stage
has a coeffcient which is of opposite sign relative to what one has in the
3-d SU(2) Higgs model. For the O(N) scalar models this has been observed by
\cite{Jak96}. It cannot be accidental but we did not pursue this phenomenon
in the present paper. The scale $\mu_3$ should not affect the results in the 
exact solution of the 3d model, but might be tuned at finite order, 
if necessary.

The finite, $\mu_3$-independent piece of the two-loop thermal mass can be 
parametrized as
$m(T)^2=k_1 g^4+k_2g^2\lambda+k_3 \lambda^2$ and the numerical values of the
coefficients are
\begin{eqnarray}
&& k_1=- 0.0390912288\nonumber\\
&& k_2=- 0.0116685842\nonumber\\
&& k_3=  0.0027102886.
\label{numcoeff}
\end{eqnarray}

\subsection{Non-static nonlocality}

The higher terms of the expansion in the external momentum squared (${\bf 
p}^2)$ of the bubble diagrams of Appendix B give rise to non-local 
interactions of the static fields. For each field a kernel can be introduced, 
what we shall denote by $\hat I^H(p^2),\hat I^G(p^2)$ and $\hat I^A(p^2)$, 
respectively. They are defined by subtracting from the full expressions of 
the corresponding bubbles the first two terms of their gradient expansions, 
which were taken into account already by 
the wave function renormalisation and the mass renormalisation locally 
(see 3.2):
\begin{equation}
\hat I^Z(p^2)=I^Z(p^2)-I^Z(0)-p^2I^{Z\prime}(0)
\end{equation}
($Z=H,G,A$). These kernels are used in the calculation of the Higgs-potential  
from the effective 3-d model at the static 1-loop level (see section 5).
Their contribution is of the form of Eq.(\ref{E1}) in Appendix E.
In place of the symbol $m$ in Eq.(\ref{E1}) 
one should use the mass of the corresponding 
static field. When working with ${\cal O}(g^4,\lambda g,\lambda^2)$ accuracy,
we should use the approximation (\ref{E7}) to the ${\bf p}$-integral.

Since $\hat I^Z(0)=0$, the first term of $(\ref{E7})$ does not contribute. This
is welcome, since this circumstance ensures that no potential linear in $\PU$ 
will be produced. Also, we notice that $\hat I^Z(p^2)=\hat I_1^Z(p^2)$ (for the 
notations, see (\ref{E2})), therefore
the integrands of the last terms are given with help of a single function.
Nonetheless, as explained in Appendix E, in the second term on the right hand 
side of (\ref{E7}) one can use the expansion of $I^Z(p^2)$ with respect to the
mass-squares of the non-static fields truncated at linear order, 
while in the third term only the mass-independent piece is to be retained. 
The  expression to be used in the first integral
we shall denote below by $\hat I^{Z1}$, while the second by $\hat I^{Z2}$.

One can point out few useful features of the integrals appearing in (\ref{E7}), 
allowing the omission of 
unimportant terms from the full expressions of the kernels. We shall discuss
these simplifying remarks first, and present only the relevant pieces from the
expressions of the kernels. 

There are two kinds of diagrams contributing to each $\hat I^Z$: 

i) the vertices are
independent of the background $\PU$, and are proportional to $g$,

ii) the vertices are proportional to $\PU$ and to $g^2, \lambda$. In this case
the two-point functions actually correspond to certain  4-point vertices, 
involving two constant external $\Phi$-lines. The corresponding
non-locality has been discussed for the $N$-vector model in \cite{Jak96}.
The way we handle them in the present paper seems to us more standardizable.

In case i) the first term of the mass expansion of $\hat I^{Z1}$ 
is going to contribute to the effective Higgs-potential a constant which will
be omitted. On the other hand  kernels of this type when multiplied by
$m^2$ in front  of the third term of (\ref{E7}) already reach the accuracy 
limiting our calculation,
 therefore their mass independent part is sufficient for the computation.

In case ii) the coupling factors in front of each diagram are already ${\cal O}
(g^4,~etc.)$, therefore they do not contribute to $\hat I^{Z2}$ in a calculation
with the present accuracy, while only their mass independent terms are to be 
used in $\hat I^{Z1}$.

Furthermore, some terms of $\hat I^Z(p^2)$ give rise in the
calculation of the effective Higgs potential to IR-finite $\bf p$-integrals 
fully factorised from the rest (this is true in particular for the subtractions 
from $I^Z(p^2)$). These contributions turn out to be proportional 
to the UV cut-off and their only role is to 
cancel against the 3-dimensional "counterterms" generated in the reduction
step (see Eq.({\ref{bare3dmass})). We need explicitly only the finite 
contributions from the integrals. There is no need to present 
those parts of the kernels  which surely do not  have any finite contribution.
 
With these remarks, we restrict ourselves to presenting only
the relevant part of the kernels for the three static fields (H,G,A), and also 
for the static ghost. 
The occurring integrals as a rule were reduced to unit numerators, 
with help of the usual "reduction" procedure \cite{Arn93}. 

The Higgs-nonlocality:
\begin{eqnarray}
\hat I^{H1}(p^2) & = &\left(-({2\over 3}\lambda^2+{15\over 16}g^4)\PU^2+
{3g^2m_G^2\over 2}-{3g^2m_a^2\over 4}\right)\int{1\over K^2(K+p)^2}\nonumber\\
&&
+3g^2p^2\left(m_G^2+{1\over 2}m_a^2+{g^2\over 8}\PU^2\right)
\int_K{1\over K^4(K+p)^2}.\nonumber\\
\hat I^{H2}(p^2) & = & -3g^2p^2\int_K\left({1\over 2K^2(K+p)^2}-{p^2\over 4K^4
(K+p)^2}\right).
\label{higgsnonl}
\end{eqnarray}

The Goldstone-nonlocality:
\begin{eqnarray}
\hat I^{G1}(p^2) & = &\left(g^2m_G^2+{1\over 2}g^2m_H^2-{3g^2\over 4}m_a^2-
{\lambda^2\PU^2\over 9}\right)\int{1\over K^2(K+p)^2}\nonumber\\
&&
+2g^2p^2\left({3m_a^2\over 4}+{m_H^2\over 2}+m_G^2\right)
\int{1\over K^4(K+p)^2},\nonumber\\
\hat I^{G2}(p^2) & = & {3\over 2}g^2p^2\int\left({1\over K^2(K+p)^2}
+{p^2\over 2K^4(K+p)^2}\right).
\label{goldstonenonl}
\end{eqnarray}

The magnetic vector nonlocality:
\begin{eqnarray}
\hat I^{A1}(p^2) &=& 8g^2m_a^2\int{1\over K^4 Q^2}\biggl[4P^2-2{(KP)^2\over K^2}
-2{(QP)^2\over Q^2}+{1\over2}\biggl(k^2+q^2\nn
&&-{(KP)^2\over P^2}-{(QP)^2\over P^2}\biggr)\biggl(2+{(KQ)^2\over K^2Q^2}\biggr)\biggr]\nn
&&+{g^2\over2}(3m_G^2+m_H^2)\int{1\over K^4 Q^2}
\biggl(k^2+q^2
-{(KP)^2\over P^2}-{(QP)^2\over P^2}\biggr)\nn
&&+ {g^4\over4}\P^2\int{1\over K^4 Q^2}\left(-2K^2
+k^2-{(KP)^2\over P^2}\right),\nn
\hat I^{A2}(p^2) &=& -4g^2\int{1\over K^2 Q^2}\biggl[4P^2-4{(KP)^2\over K^2}
+\left(k^2-{(KP)^2\over P^2}\right)\left({(KQ)^2\over K^2Q^2}-1\right)\nn
&&+3\left(k^2-{(KP)^2\over P^2}\right){K^2-Q^2\over K^2}\biggr].
\label{vectnonloc}
\end{eqnarray}

The ghost-nonlocality (no ghost contribution arises to the second integral of
(\ref{E7})):
\begin{equation}
\hat I^{C1}(p^2)={1\over 2}g^2m_a^2\int {1\over K^2(K+p)^2}-g^2m_a^2p^2
\int{1\over K^4(K+p)^2}.
\label{ghostnonloc}
\end{equation}
Two useful remarks can be made concerning the magnetic vector nonlocality:

In the second equation of (\ref{vectnonloc}) sometimes only combinations of 
terms in the integrands prove to be proportional to $p^2$.
Therefore the corresponding weighted integrals should not be performed term 
by term.

There are parts of the kernel (\ref{vectnonloc}) 
which could not be reduced to unit
numerator, but are proportional to powers of ${\bf k}^2$. We shall see that
upon calculating their contribution to the Higgs effective potential
from the 3-d gauge-Higgs effective system, they combine with non-local 
contributions
to the potential from the static $A_0$ integration (see the end of subsection 
4.1 below) into formally Lorentz-invariant  contributions.

\section{Static electric fluctuations}

\subsection{Contribution to the potential term}

The $A_0$-integration resummed with help of a thermal mass $m_D$ yields
at 1-loop:
\begin{equation}
V^{(1-loop)}={3\over2}I_3(m_{A0}),
\end{equation}
defined through the integral
\begin{equation}
I_3(m)=T\int_{\bf k}\ln ({\bf k}^2+m^2)
\end{equation}
(its value appears in Appendix A).

At two loops, vacuum diagrams involving one $A_0$ propagator are listed with
their analytic definitions in Appendix D. In addition also three counterterm
contributions of (\ref{Lct}) should be taken into account on 1-loop level.
They can be expressed (up to constant terms) with help of $I^\prime_3(m_{A0})$
where the prime denotes differentiation with respect to $m_{A0}^2$:
\begin{equation}
V_{CT}^{(A0)}={3T\over 2}I^\prime_3(m_{A0})
\left[-m_D^2-Z_{A1}m_{A0}^2
+{g^2\PU^2\over 4}(Z_{A1}+Z_{\Phi1}+2Z_{g1})\right].
\end{equation}
When only terms up to ${\cal O}(g^4)$ are retained and further constants are
thrown away, one finds the contribution
\begin{equation}
V_{CT}^{(A0)}={3Tm_D^2m_{A0}\over 8\pi}+
{3g^2T\PU^2\Lambda \over 16\pi^2}(Z_{\Phi1}+2Z_{g1}).
\end{equation}

The evaluation of 2-loop diagrams with topology of 'Figure 8' does not pose 
any problem, since their expressions are automatically factorised into the 
product of static and nonstatic (sum-)integrals. The evaluation scheme of 
'setting sun' diagrams needs, however, some explanation. 

Their general form is as appears in (\ref{E1}). Exploiting the analysis of
Appendix E we see that the final result of the $\bf p$-integration
depends non-analytically on $m_{A0}^2$. The non-analytic piece comes from the
first term of the Taylor expansion of the kernel $I$ with respect 
to ${\bf p}^2$: $I(0)\equiv I_3^\prime (m_{A0})$. 
In other words, this contribution is the product of a 
static and of a nonstatic integral, similarly to the 'Figure 8' topology. 
In summary, we find that contributions linear in $m_{A0}$ come from the 
thermal counterterm and from the "factorised" 2-loop expressions. They sum up 
to
\begin{equation}
{3m_{A0}\over 8\pi}\left(-{5g^2\over 2\pi^2}\Lambda T+{5g^2\over 6}T^2-m_D^2
\right).
\end{equation}

An optimal choice for the resummation mass ($m_D$) is when it is not 
receiving any finite contribution from higher orders of the perturbation theory.
This requirement leads to the equality
\begin{equation}
m_D^2={5\over 6}g^2T^2,
\end{equation}
which means that the two-loop $A_0$-contribution at the level of the "local
approximation" yields exclusively a linearly diverging, non-polynomial 
contribution to the potential energy term of the effective 3-d gauge-Higgs 
model:
\begin{equation}
V^{(2-loop)}_{loc}={15g^2\over 16\pi^3}m_{A0}\Lambda T.
\label{nonpoldiv}
\end{equation}
This term plays essential role in the consistent 2-loop perturbative solution
of the effective model (see Section 5).

Further terms of the expansion of $I(p^2)$ are interpreted as  
 higher derivative kinetic terms of $A_0$.
Their evaluation is based on (\ref{E7}). The only difference relative to the 
content of subsection 3.4 is that the $A_0$ integration is performed already
at this stage, therefore the contribution from its non-local kinetic
action modifies now the potential energy of the Higgs field. When the terms
yielding ${\cal O}(g^4,g^2\lambda ,\lambda^2)$ accuracy are retained, 
it turns out that they actually contribute only to the mass term:
\begin{eqnarray}
&&\!\!\Phi^\dagger\Phi\,\Biggl\{-g^2\left({3g^2\over 2}
+{3\over8}\lambda\right)
\int{1\over p^2}\int{1\over Q^4}-{3g^4\over 4}\int{1\over p^2}
\int{Q_0^2\over Q^6}\nonumber\\
&&+3g^4\int{1\over p^2Q^2K^4}\left(2p^2-{(kp)^2\over K^2}-
{(qp)^2\over Q^2}+2Q_0^2+Q_0^2{(KQ)^2\over K^2Q^2}\right)\nonumber\\
&&+{3\over2}g^2\left(\lambda +{g^2\over4}\right)
\int{Q_0^2\over p^2Q^2K^4}-{3\over8}g^4\int{1\over p^2Q^2K^2}\nonumber\\
&&+{3g^4\over 2}\int{1\over p^4Q^2K^2}\left(2p^2-2{(kp)^2\over K^2}+
Q_0^2{(KQ)^2\over K^2Q^2}-Q_0^2\right)\nonumber\\
&&+{9g^4\over 2}\int{Q_0^2(K^2-Q^2)\over p^4Q^2K^4}\Biggr\}.
\label{a0mass}
\end{eqnarray}

In view of the remark at the end of subsection 3.4 
on non-static nonlocalities, it is not convenient to evaluate 
explicitly this contribution already at the present stage.
Its Lorentz non-invariant pieces are going to combine in the expression of 
the final effective 
Higgs potential with contributions from the static magnetic vector fluctuations
 into  simpler Lorentz invariant integrals (see section 5).

\subsection{$A_0$ contribution to the gauge-Higgs kinetic terms}

The $A_0$-bubble contributes to the self-energy of the magnetic vector
and of the Higgs (SU(2) singlet) scalar fluctuations. The corresponding
analytic expressions are given in Appendix B.

There are two ways to treat
these contributions. The first is to apply the gradient expansion again,
and retain the finite mass correction and the wave function rescaling
factor from the infinite series. This is the approach we followed in case of 
non-static fluctuations. Then the magnetic vector receives only
field rescaling correction:
\begin{equation}
\delta Z_{(A0)}^A={g^2T\over 24\pi m_{A0}}.
\label{ascale}
\end{equation}
The Higgs particle receives both finite mass correction and field rescaling
factor from the $A_0$-bubble:
\begin{eqnarray}
\delta m_{(A0)}^H & = & -{3g^4\PU^2T\over 64\pi m_{A0}},\nonumber\\
\delta Z_{(A0)}^H & = & {g^4\PU^2\over 512\pi m_{A0}^3}.
\label{rescal}
\end{eqnarray}
One should
note that the rescaling factors are ${\cal O}(g)$ what is just the order
one has to keep to have the ${\cal O}(g^4)$ accurate effective action
\cite{Bod94}.  The different behavior of the Higgs and of the Goldstone fields
reflects the breakdown of the gauge symmetry in presence of $\PU$.

The Higgs-field mass and field rescaling corrections appear only in
a non-zero $\PU$ background. The $\PU$-dependence of 
these quantities will be treated as a non-fluctuating {\it parametric} 
effect in the course of solving the effective gauge-Higgs model. 

In this case the
kernel of the one loop integral is not analytic in $p^2$, therefore no
basis can be provided for the gradient expansion.
Therefore it is important to proceed also the other way, 
that is to keep the original bubble as a whole in form of a nonlocal part of the
effective action. Since its coefficient is $g^4$, in the perturbative
solution of the effective model it is sufficient to compute its contribution
to 1-loop accuracy. (This amounts actually to a two-step evaluation of the
of the purely static AAH and AAV setting sun diagrams.) 

The difference between the two approaches from the point of view of the
quantitative characterisation
of the phase transition will be discussed within the perturbative framework
in the next Section.

\section{Two-loop effective Higgs potential from the 3-d effective theory}

In this and the next sections we are going to compare various versions of the 
3-d theory by calculating perturbatively with their help the characterisation 
of the electroweak phase transition. This we achieve by finding the respective
point of degeneracy of the effective Higgs potential in each approximation.
The analysis will be performed for $g=2/3$ and $M_W=80.6$GeV, and for 
various choices of the $T=0$ Higgs mass. (The scheme of the determination of 
the renormalised parameters is sketched in Appendix C).

\subsection{Approximations to the effective theory}

We start by summarising the effective Lagrangian obtained from the calculations
of sections 3 and 4. The parameters were calculated with accuracy ${\cal O}
(g^4,\lambda g^2, \lambda^2)$:
\begin{equation}
{\cal L}_{3D}={1\over 4} F_{ij}^aF_{ij}^a
+ {1\over 2} (\nabla_i\Phi)^{\dagger}(\nabla_i\Phi)
+ {1\over 2} m(T)^2 \Phi^\dagger\Phi + {\lambda_3\over24}(\Phi^\dagger\Phi)^2
- {1\over4\pi}m_{A0}^3+ {\cal L}_{nonloc}+{\cal L}_{CT},
\label{eq:3dact}
\end{equation}
where
\begin{eqnarray}
& F_{ij}^a & =\partial_i A_n^a -\partial_j A_m^a + g_3 
\epsilon_{abc}A_i^aA_j^b \nonumber\\
& \nabla_i & =\partial_i - ig_3A_i^a\tau_a,\nonumber\\
& m_{A0}^2 & =g_3^2\left({5\over6}T+{1\over4}\Phi^\dagger\Phi\right)
\nonumber\\
& m(T)^2    & =m^2 + T^2\biggl[{3\over16}g^2+{\lambda\over12}
+g^4\left({81\over32}I_{20}+{81\over16}K_{00}+{157\over7680\pi^2}\right)
\nonumber\\
&&
+g^2\lambda\left({3\over4}I_{20}+{3\over2}K_{00}+{1\over192\pi^2}\right)
\nonumber\\
&&-\lambda^2\left({1\over6}I_{20}+{1\over3}K_{00}\right)
+\Delta_{A0}^{nonloc}\nn
&&+{1\over 8\pi^2}\bigl({81g^4\over 16}+{3g^2\lambda\over2}-{\lambda^2\over3}
\bigr)\ln{\mu_3\over T}\biggr].
\end{eqnarray}         
The dependence of the thermal mass on the 3-d scale is incorrect. The only 
way to ensure the correct scale dependence of the thermal mass is to include 
also into the local approximations the effect of the non-static and and some
of the static nonlocalities. 

For this we write down their contribution to the effective Higgs potential
combined
from (\ref{higgsnonl}), (\ref{goldstonenonl}) (\ref{ghostnonloc}) on one hand, 
and from (\ref{vectnonloc}) and (\ref{a0mass}) on the other. The contributions 
from the different nonlocalities are
expressed through some integrals appearing in Appendix A:

{\it Higgs+Goldstone}:
\begin{equation}
{1\over 2}\PU^2\left[H_{43}(0,0,0)\left(-\lambda^2
-{27\over 16}g^4+3\lambda g^2\right)-{\displaystyle{d\over dm_1^2}C(0,0)}
\left({15\over 8}g^4+{9\over 4}\lambda g^2\right)\right]
\label{higonl}
\end{equation}

{\it Magnetic and electric vector:}
\begin{eqnarray}
&{1\over 2}\PU^2 \Biggl[H_{43}(0,0,0)\left({129\over 8}g^4+
{3\over 2}\lambda g^2\right)+{\displaystyle{d\over dm_1^2}C(0,0)}
\left( {81\over 16}g^4-{3\over 4}\lambda g^2\right)\nn
&
-{3\over 2}g^4\displaystyle{{d\over dm_1^2}}L_4(0,0)\Biggr].
\label{magenl}
\end{eqnarray}
{\it Ghost}:
\begin{equation}
{1\over2}\PU^2\left[{3\over4}g^4H_{43}(0,0,0)+{3\over2}g^4{d\over dm_1^2}C(0,0)
\right].
\end{equation}
Using the features of the 3 basic integrals ($H_{43}, C, L_4$),
listed in Appendix A, the final numerical expression for the sum
of the local (e.g. (\ref{numcoeff}))
plus non-local ${\cal O}(g^4)$  thermal mass terms is the following:
\begin{eqnarray}
&
{1\over 2}\PU^2\biggl(0.0918423g_3^4+0.0314655\lambda_3 g_3^2
-0.0047642\lambda_3^2\nn
&
+{1\over16\pi^2}\ln{\mu_3\over T}\left({1\over3}\lambda^2-{3\over2}
\lambda g^2-{81\over16}g^4\right)\biggr).
\end{eqnarray}

The 2-loop contribution to $m^2(T)$ could be much influenced
by the choice of the 3d renormalisation scale. Through the 4-d 
renormalisation of the mass-parameter $m^2$ it is sensitive also to the 4-d 
normalisation scale (see Appendix C).   

All fields in (\ref{eq:3dact}) are renormalised 4-d fields multiplied by 
$\sqrt{T}$. The 3-d gauge coupling in the kinetic terms is simply 
$g_3=g\sqrt{T}$, due to the fact that all finite T-independent corrections are
included into $Z_g$ (cf. eq.(\ref{zcoupling})). Also the quartic coupling 
takes the simple form $\lambda_3=\lambda T$, in view of the particular 
renormalisation condition (\ref{rencond}). 

The cut-off dependent part of the mass term is hidden in ${\cal L}_{CT}$.
It ensures the cancellation of the cut-off dependence
of the 3-d calculations. Except for the non-polynomial term, we are not going
to discuss this cancellation. The formal correctness of our effective Higgs 
potential will be verified by checking its agreement with results of
other 3-d calculations, expressed through the parameters 
$m^2(T),g_3^2,\lambda_3, m_H^2,m_G^2,m_a^2$. 
Off course these parameters themselves are connected to the 4-d couplings 
differently.

The non-polynomial term of the above Lagrangian is simply the result of the 
1-loop $A_0$-integration. 

Terms of the  remaining non-local Lagrangian 
${\cal L}_{nonloc}$ are all of the generic form
\begin{equation}
{\cal L}_{nonloc}= {1\over2}\phi(-p){\cal N}(p)\phi(p)
\end{equation}
($\phi (p)$ refers to general 3-d fields). The kernels ${\cal N}(p)$ are the 
result of purely static $A_0$-loops, listed in Appendix B.
\begin{eqnarray}
&
{\cal N}_H(p)=-{3g_3^4\over8}\PU^2\int_{\bf k}\displaystyle{{1\over 
(k^2+m_{A0}^2)((k+p)^2+m_{A0}^2)}}\nn
&
{\cal N}_{ij}^a(p)=-4g^2\int_{\bf k}\displaystyle{\biggl
[{k_s\Pi_{si}k_r\Pi_{rj}\over (k^2+m_{A0}^2)
((k+p)^2+m_{A0}^2)}-{1\over3}{k^2\delta_{ij}\over (k^2+m_{A0}^2)^2}}\biggr]
\end{eqnarray}

In order to see clearly the effect of the non-polynomial and nonlocal terms we
shall consider the effective potential in three different approximations.
\vskip .3truecm
{\it i) Local, polynomial approximation}

In this approximation non-local effects due  
to the $A_0$-bubbles are neglected. The parametric "susceptibilities"
of the $A_i$ and $\Phi_H$ fields arising from the static $A_0$-bubble
are included into the approximation (the background $\Phi_0$ appearing in 
their expressions is a non-fluctuating parameter!). 
The non-polynomial term is expanded to 
fourth power around $m_{D}^3$. In presence of non-zero Higgs background the 
Lagrangean actually breaks(!) the gauge symmetry (some coefficients explicitly
depend on the value of the background field, cf. (\ref{rescal})). 
The approximate form of the effective Lagrangian is:
\begin{eqnarray}
{\cal L}_{LP} &=& {1\over 4\mu}F_{ij}^aF_{ij}^a+{1\over 2}(\nabla_i\Phi )
^\dagger(\nabla_i\Phi )+{1\over 2}Z_H(\partial_i\xi_H)^2\nonumber\\
&&
+V_3(\Phi^\dagger\Phi )+ig_3\epsilon_{abc}p_ic^\dagger_a(p)A_i^a(q)c_c(k),
\end{eqnarray}
with the  local potential
\begin{equation}
V_3(\Phi^\dagger\Phi )={1\over 2}(m(T)^2+\delta m^2)\Phi^\dagger\Phi +{1\over 2}
\delta m_H^2\xi_H^2+{\lambda_3 +\delta\lambda\over 24}(\Phi^\dagger\Phi )^2,
\end{equation}
where 
\begin{eqnarray}
&
\mu = (1+{\displaystyle g_3^2\over \displaystyle 24\pi m_{A0}})^{-1},
&\delta\lambda =-{9g_3^4\over 64\pi m_D}\nonumber\\
&
 \delta m^2 = -{\displaystyle 3g_3^2m_D\over \displaystyle 16\pi},
&\delta m_H^2 = -{3g_3^4\PU^2\over
64\pi m_{A0}},\nonumber\\
&
Z_H ={\displaystyle 3g_3^4\PU^2\over \displaystyle 512\pi m_{A0}^3},&
\end{eqnarray}
(the expressions of $m(T)^2,~g_3^2,~\lambda_3$ have been given above). 
We note, that also a term $Z_H(A_i^a)^2\xi_H^2/2$ is present, but it would 
contribute to higher orders in the couplings, and such terms will not be 
displayed here. 

In order to apply perturbation theory with canonical kinetic terms, one has 
to rescale both the $A_i$ and $\xi_H$ fields:
\begin{equation}
\bar g_3=g_3\mu^{1/2},~~~\bar A_i=A_i\mu^{-1/2},~~~\bar\xi =\xi_H(1+Z_H/2).
\end{equation}
Below we continue to use the {\it un}barred notation for these quantities!

After rescalings and the $\PU$-shift in the $R_\alpha$-gauge the action 
density is written as
\begin{eqnarray}
{\cal L}_{LP} &=& V_3((1-Z_H)\PU^2)\nonumber\\
&&+{1\over 2}A_i^a(-k)\left[(k^2+m_a^2)\delta_{ij}-(1-{1\over \alpha})k_ik_j
\right]A_i^a(k)+c_a^\dagger k^2c_a(k)\nonumber\\
&&+{1\over 2}\xi_H(-k)[k^2+m_h^2]\xi_H(k)+{1\over 2}\xi_a(-k)[k^2+m_g^2]\xi_a(k)
\nonumber\\
&&+ig_3S_{abc}^{ijk}(p,q,k)A_i^a(p)A_j^b(q)A_k^c(k)+{1\over 8}A_i^2[g_3^2\xi_a^2
+g_{h}^2\xi_H^2]\nonumber\\
&&+{g_3^2\over 4}[(A_i^aA_i^a)^2-(A_i^aA_j^a)^2]+{g_{h}^2\PU \over 4}
A_i^2\xi_H\nonumber\\
&&+ig_3\epsilon_{abc}p_ic_a^\dagger (p)A_i^a(q)c_c(k)\nonumber\\
&& -i{1\over 2}A_i^a(p)[g_{h}(k-q)_i\xi_a(q)\xi_H(k)+
g_3\epsilon_{abc}k_i\xi_b(q)\xi_c(k)]\nonumber\\
&&
+{1\over 24}(\lambda_{HH}\xi_H^4+2\lambda_{HG}\xi_H^2\xi_G^2+\lambda_{GG}
\xi_a^2)\nonumber\\
&&\PU{1 \over 6}\xi_H(q_{HHH}\xi_H^2+q_{HGG}\xi_a^2) +{\cal L}_{CT}^{polyn}+
{15g^2\over 16\pi^3}m_{A0}\Lambda.
\label{3dnonloclagr}
\end{eqnarray}
Here the expressions for the different couplings read as
\begin{eqnarray}
 g_{h}^2=g_3^2(1-Z_H), &
\lambda_{GG}=\lambda_3+\delta\lambda\nonumber\\
\lambda_{HH}=(\lambda_3+\delta\lambda )(1-2Z_H), & \lambda_{HG}=
(\lambda_3+\delta\lambda )(1-Z_H),\nonumber\\
q_{HHH}=(\lambda_3+\delta\lambda )(1-2Z_H), & q_{HGG}=(\lambda_3+\delta\lambda)
(1-Z_H).
\end{eqnarray}
 The effective masses of the different fields are
\begin{eqnarray}
m_a^2 &=& {g_3^2\over 4}\PU^2(1-Z_H),\nonumber\\
m_h^2 &=& (m(T)^2+\delta m^2+\delta m_H^2)(1-Z_H)+
{\lambda_3+\delta\lambda\over 2}\PU^2(1-2Z_H),\nonumber\\
m_g^2 &=& m(T)^2+\delta m^2+{\lambda_3+\delta\lambda\over 6}\PU^2(1-Z_H).
\end{eqnarray}
\vskip .3truecm
{\it ii) Local, non-polynomial approximation}

The rescaling and the shift of the fields proceeds as before. The main
difference is that here the order of expanding and shifting the $\Phi$-field
in the non-polynomial term is changed relative to case i).
The form of the Lagrangian can be equally written in the form 
(\ref{3dnonloclagr}). Some expressions and
 constants appearing in it are modified relative to case i):
\begin{equation}
V_3(\Phi^\dagger\Phi )={1\over 2}(m(T)^2+\delta m^2)\Phi^\dagger\Phi +{1\over 2}
\delta m_H^2\xi_H^2+{\lambda_3 +\delta\lambda\over 24}(\Phi^\dagger\Phi )^2
-{1\over 4\pi}m_{A0}^3,
\end{equation}

\begin{eqnarray}
m_{A0}^2 &=& m_D^2+{1\over 4}g_{h}^2\PU^2, \nonumber\\
\delta m^2 &=& -{3g_3^2 m_{A0}\over 16\pi}, \nonumber\\
m_h^2 &=& (m(T)^2+\delta m^2+\delta m_H^2)(1-Z_H)+{\lambda_3\over 2}\PU^2
(1-2Z_H),\nonumber\\
m_g^2 &=& m(T)^2+\delta m^2+{\lambda_3\over 6}\PU^2(1-Z_H),
\end{eqnarray}

\begin{eqnarray}
&q_{HGG}=\lambda_3(1-Z_H)-{\displaystyle{9g_{h}^2g_3^2\over 64\pi m_{A0}}},\nn
&q_{HHH}=q_{HGG}(1-Z_H)+{\displaystyle{3g_{h}^6\PU^2\over 256 m_{A0}^3}},\nn
&\lambda_{GG}=\lambda_3-{\displaystyle{9g_{h}^4\over 64\pi m_{A0}}},\nn 
&\lambda_{HG}=\lambda_{GG}(1-Z_H)+
{\displaystyle{9g_{h}^4g_3^2\PU^2\over 64\pi m_{A0}^3}},
\nonumber\\
&\lambda_{HH}=\lambda_{HG}(1-Z_H)+
\displaystyle{{9g_{h}^6\PU^2\over 256\pi m_{A0}^3}
-{9g_{h}^8\PU^4\over 1024\pi m_{A0}^5}}. 
\end{eqnarray}

{\it iii) Nonlocal, nonpolynomial representation}

Its Lagrangean density coincides with (\ref{eq:3dact}), and the major issue 
of our investigation is the quantitative comparison of its ${\cal O}(g^4)$
solution to the approximate solutions corresponding to cases i) and ii).
The couplings $q_{HHH},q_{HGG}$ and $\lambda_{HG},\lambda_{GG},\lambda_{HH}$
agree with their expression derived for case ii) 
when one puts $Z_H=0, \delta m_H^2=0$.

\subsection{Variations on the effective Higgs potential}

The 1-loop correction to the "classical" value of $V_3(\PU^2)$ has the unique 
form for all approximations:
\begin{equation}
V_{eff}^{(1)}=-{1\over 12\pi}(6m_a^3+m_h^3+3m_g^3)+{15g_3^2\over 
16\pi^3}m_{A0}\Lambda+V_{CT}^{polin}(\PU ).
\end{equation}
In this expression cut-off dependent terms with polynomial dependence on 
$\PU$ are not displayed, their cancellation is taken for granted. Though 
the form of this expression is unique, one should keep in mind the variable
meaning of the different masses from one approximation to the other. 
From the point  of view of consistent 
cancellation of non-polynomial divergences it is important to notice, that
the 1-loop linear divergences of the Higgs- and Goldstone-fields by 
the contribution of the shifted non-polynomial term  to their respective mass
produce a non-polynomial cut-off dependent term 
$(-3g_3^2/16\pi^3)\Lambda m_{A0}$ (in cases ii) and iii) only)
to be compared with
the induced non-polynomial divergence eq.(\ref{nonpoldiv}). In the first two
approximation schemes no further divergences of this form are produced,
therefore their cut-off dependence is clearly different from what is generated
in the course of the integration. Below, for cases i) and ii) we assume the 
presence of the appropriately modified counterterms.

The 2-loop diagrams of a local, 3-d gauge-Higgs system are the same as in 
4-d. Just the functions appearing in the expressions are defined as 
three-dimensional 
momentum integrals. In purely algebraic steps one can work out the 
general representation given in terms of the functions $I_3(m^2), 
H_3(m_1,m_2,m_3)$ (see Appendix A), to arrive finally at the following 
2-loop contribution:
\begin{eqnarray}
V_{eff}^{(2)} &=&
L_0\biggl( {63 g^2 m_a^2\over8} - {3 g_h^2 m_a^2\over2}
 + {3 g^2 m_G^2\over2}+ {3 g_h^2 m_G^2\over4}
+ {3 g^2 m_H^2\over4}\nn
&& - {q_{GGH}^2+q_{HHH}^2\over12}\PU^2\biggr)
+{1\over128 \pi^2}(5\lambda_{GG} m_G^2+2\lambda_{GH} m_G m_H
+\lambda_{HH} m_H^2)\nn
&&+{g^2\over128 \pi^2}\biggl(
 (9-63\ln3)m_a^2- {3g_h^2m_a^2\over g^2} + 12 m_a m_G
- {3g_h^2m_am_G\over g^2}\nn
&& {6 g_h^2 m_a m_H\over g^2}
+3 m_G^2 + {3 g_h^2 m_G m_H\over g^2} + {3g_h^2 m_H^2\over2g^2}\biggr)\nn
&&-{3 g_h^2\over 128 \pi^2} {m_H-m_G\over m_a} ( m_H^2-m_G^2 )\nn
&&+ {q_{GGH}^2\PU^2\over192 \pi^2} \ln{2 m_G + m_H\over T}
+ {q_{HHH}^2\PU^2\over192 \pi^2} \ln{3 m_H\over T}\nn
&&+ {3\over64 \pi^2}\Biggl[g_h^2 m_H^2 \ln{m_a+m_H\over2m_a+m_H}
 + {g_h^2 m_H^4\over 4 m_a^2} \ln{m_H(2m_a+m_H)\over(m_a+m_H)^2}\nn
&& +  {g_h^2(m_H^2-m_G^2)^2\over 2 m_a^2} \ln{m_a+m_G+m_H\over m_G+m_H}\nn
&& - g_h^2\left(m_H^2+m_G^2-{m_a^2\over2}\right) \ln{m_a+m_G+m_H\over T}\nn
&&+  g^2\left( {m_a^2\over2}-2m_G^2\right) \ln{m_a+2m_G\over T}
 -{g_h^2 m_a^2\over2} \ln{m_a+m_H\over T}\nn
&& +2g_h^2 m_a^2 \ln{2m_a+m_H\over T}
  - 11 m_a^2 \ln{m_a\over T}\Biggr].
\label{pot2loop}
\end{eqnarray}
The final step of composing the full effective potential corresponds to 
picking up the contributions of the purely static {\bf VAA}, and {\bf HAA} 
diagrams:
\begin{eqnarray}
V_{eff}^{(2nonloc)} &=& 6g_3^2L_0(m_{A0}^2-{3\over 8}m_a^2)+
{3g_3^2\over 32\pi^2}(m_{A0}^2+
2m_{A0}m_a)-{12g_3^2\over 16\pi^3}\Lambda m_{A0}\nonumber\\
&&-{3g_3^2\over 32\pi^2}\bigl[(4m_{A0}^2-m_a^2)\ln{2m_{A0}+m_a\over \mu_3}\nn
&& -{1\over 2}m_a^2\ln{2m_{A0}+m_H\over \mu_3}\bigr].
\label{pota0nonloc}
\end{eqnarray}                                    
The cancellation of the linear divergence, nonpolynomial in $\PU^2$ can be
seen explicitly.

The 2-loop corrections to the effective potential, given by 
Eqs.(\ref{pot2loop})
and (\ref{pota0nonloc}) can be compared to the results of the direct 4-d
calculation of \cite{Fod94,Buc95}. Those calculations were done with
different regularisation and renormalisation scheme, therefore in the 
expression
of the potential terms reflecting this difference are seen. Still, using
the leading expressions for our couplings $g_h,q_{GGH},q_{HHH},
\lambda_{GG},\lambda_{GH}$ and $\lambda_{HH}$, all logarithmic terms expressed
through the propagator masses agree. Also our polynomial terms, except
those proportional to the regularisation dependent coefficient $L_0$, have
their exact counterpart in the 4-d expression. The $T$-independent contribution
appearing in the 4d result has no corresponding term in our case, an obvious
effect of the difference of the renormalisation schemes. The polynomial
terms explicitly depending on $\bar \mu$ and proportional to the constants
$c_1,c_2$ characteristic for the dimensional regularisation, can be
compared to our terms proportional to $L_0$. Choosing in both potentials
$\mu =T$, one finds in the 4-d expression for the coefficients of the terms 
$g^2m_W^2$, $g^2(m_H^2+3m_G^2)$ and $\lambda^2\PU^2$, the values
$2.441\times 10^{-2},4.012\times 10^{-3}, -2.239\times 10^{-4}$, respectively.
The corresponding coefficients from our expression are: $2.765\times 10^{-2},
5.027\times 10^{-3}, -5.374\times 10^{-4}$. Therefore the only origin of
considerable deviations in physical quantities could be the
effect of the reduction on various couplings. Our numerical experience was 
that the ${\cal O}(g^4)T^2$ correction in $m^2(T)$ accounts for essentially all
differences. 

\section{Phase transition characteristics}

In this section we describe and discuss the phase transition in the 
three perturbative approximations introduced in the previous section. 
Our analysis will cover the Higgs mass range 30 to 120 GeV. Our main interest 
lies in finding the percentual variation in the following physical quantities:
the critical temperature ($T_c$), the Higgs discontinuity ($\Phi_c$), 
the surface tension ($\sigma$) and the latent heat ($L_c$). The amount of 
variation from one approximation to the other will give an idea of the
theoretical error coming from the reduction step. 

The first step of the calculation consists to choose values for
$g$ and $\lambda$. For example one can fix $g=2/3$ and tune $\lambda$ 
guided by  the tree-level Higgs-to-gauge boson mass ratio. 
All dimensional quantities are scaled by appropriate 
powers of $|m|$ (what practically amounts to set $|m|=1$ in the
expressions of the Higgs effective potential). 

Next, one finds from the degeneracy condition of the effective potential
the ratio $T_c/|m|$. Here we have to discuss a phenomenon already noticed by
\cite{Fod94}. It has been observed that for Higgs mass values $m_H\geq 100$GeV
when the temperature is lowered one reaches the barrier temperature before
the degenaracy of the minima would occur. The phenomenon was traced back to
the action of the term $\sim g^2m_{A0}(m_H+3m_G)$ in the effective potential. 
In our non-polynomial approximation the $A_0$-integration contributes
a negative term to $m^2(T)$ (i.e. $\delta m^2$), which acts even stronger, 
and the same phenomenon is generated, when using $\mu_3 =T$,  already for 
$m_H\leq 70$GeV. However, by choosing somewhat more exotic value for the
normalisation scale ($\mu_3 =T/17$), we could follow the transition 
up to $m_H=120$GeV. We have used this value in the whole temperature range. 
It has been checked for $m_H=35$GeV that the variation of $\mu_3$ leads in 
the physical quantities to negligible changes.

Finally, the relations of Appendix C allow to express with ${\cal O}(g^4)$
accuracy the ratios $M_W/|m|$ and $M_H/|m|$  with help of $g,\lambda$ and
$T_c/|m|$ (the dependence on the latter appears through our choice of the 4-d 
normalisation point). The pole mass $M_H$ resulting from this calculation
appears in column 2 of Table 1, where $M_W=80.6$GeV sets the scale in physical
units. Our 4-d renormalisation scheme leads to somewhat smaller mass shifts, 
than the scheme used in \cite{Far94}.

In order to present physically meaningful plots one has to eliminate 
from these relations $|m|$ and scale everything by an appropriate physical
quantity. 
In Fig. 1 (coulumn 3 of Table 1) we present $T_c/M_H$ as a function of 
$M_H/M_W$ in the non-polynomial
non-local (NNA) approximation (upper case notation for the masses always refer
to T=0 pole mass). 
Internally,
our different approximations do not affect the critical temperature, they all 
agree within better than 1\%, as can be seen in Table 2.
The curve in Fig.1 is, however, systematically  below the 
data appearing in Table 2 of \cite{Far94}.  15\% deviation is observed
for $m_H=$35GeV, which gradually decreases to 8\% for $m_H=90$GeV.
 If one wishes to compare to the 2-loop 4d calculations of 
\cite{Fod94,Buc95} one has to use their coupling point: $g=0.63722, m_W=80.22$.
Again we find that our $T_c/m_H$ values are 10\% lower than those appearing
in Fig.4 of \cite{Buc95}. Our ${\cal O}(g^4)T^2$ correction to $m^2(T)$ is
about 10\% of the 1-loop value (and is about 9 times larger than what was 
found in \cite{Far94}), therefore one qualitatively understands that the
barrier temperature is brought down in about the same proportion. At least in
the region $M_W\simeq M_H$ the transition is already so weak that the
transition temperature should agree very well with the temperatures limiting
the metastability range.

In Fig. 2  (column 4 of Table 1) the order parameter discontinuity is 
displayed in proportion 
to $T_c$. Here the agreement is extremely good in the whole 35-90 GeV range 
both in comparison to \cite{Far94,Kaj96} and \cite{Buc95}. Also the variance
of our different approximations (see Table 2) is minimal. 

The most interesting is the case of the surface tension, 
which is shown for all three approximations in Fig.4 in $(GeV)^4$ units 
(the dimensionless ratio appears in column 5 of Table 1). 
We did not observe any strengthening tendency for larger Higgs masses, in 
contrast to \cite{Fod94}. This systematic difference seems to be 
correlated with the extended range where we find phase transition. 
It leads to 
$\sigma_c$ values in the range $M_H=50-80$GeV which perfectly agree with
the perturbative values quoted by \cite{Kaj96}.
The dispersion between the values of our different approximations is much 
larger in the high mass range (Table 2) than for other quantities, reflecting the increased 
sensitivity to non-local and non-polynomial contributions. 

The situation is just a little less satisfactory in
case of the latent heat (Fig.3 and column 6 of Table 1).
Our approximations are 10-15\% above $L_c/T_c^4$ curve of Fig.6 of 
\cite{Buc95} and also of the perturbative values in Table 6 of \cite{Kaj96}.

\section{Conclusions}
In this paper we have attempted to discuss in great detail the reduction
strategy, allowing non-renormalisable and non-local effective interactions.
We have made explicit the effect of various approximations on the
phase transition characteristics of the SU(2) Higgs model. Our investigation
remained within the framework of the perturbation theory. By comparing
with other perturbative studies \cite{Fod94,Buc95,Far94} we have pointed out 
that the $T_c/M_H$ ratio is quite sensitive to the choice of the 
3-d renormalisation scale. The order parameter discontinuity and the 
the surface tension were found only moderately sensitive,
while the latent heat in the $\overline{MS}$ scheme seems to drop faster with
increasing Higgs mass. The minimum of the surface tension when $M_H$ is varied 
has disappeared. One might wonder to what extent is the strengthening
effect observed in the 4-d perturbative treatment with increasing 
Higgs mass physical at all. 

The local polynomial and the local non-polynomial approximations start to show 
important (larger than 5\%) deviations from the nonlocal, nonpolynomial
version of the effective model only for $M_H\geq 80$GeV (also below
30GeV). Since in these  
regions the perturbation theory becomes anyhow unreliable, we can say that
the application of the reduction to the description of the electroweak phase 
transition in the relevant Higgs mass range ($M_H\sim M_W$) 
could be as accurate as 5\%. This can be true for dimensionless ratios, but 
not for $T_c/M_H$.

The present assessment of the accuracy is not universal, though the structure 
of the analysis is quite well fitting  other field theoretic models, too. 
For instance, our methods could be applied to extended versions
of the electroweak theory \cite{Los96,Cli96,Lai96} which are accessible to
quantitative non-perturbative investigations only in 3-d reduced form. 

One knows that non-perturbative studies of the SU(2) Higgs transition
led to lower $T_c/M_H$ ratio, than the perturbative prediction \cite{Fod95}, 
and at least for $M_H=80$GeV the surface tension is much lower than it
was thought on the basis of the strenghtening effect \cite{Csi96}. 
The results from the PPI-reduction seems to push the perturbative 
phase transition characteristics towards this direction.
The expressions of the mass parameter of the effective theory 
(\ref{bare3dmass})  and the T=0 pole masses (Appendix C) provide the necessary
background for the analysis of 3-d non-perturbative investigations following
the strategy advocated by \cite{Kaj93,Kaj96}. This will be the subject of 
a forthcoming publication \cite{KarTo}.

\appendix

\section{Important integrals}

All integrals have been performed with 3-d momentum cut-off $\Lambda$. The 
capital subscript denotes finite temperature sum-integral, the bold 
lower case index refers to 3-d momentum space integration. The prime
on the sum-integral means summation over the non-static modes only.

1.
\begin{eqnarray}
I_4(m) &=& I_0+I_1m^2+I_2m^4+...\nn
I_1 &=& {\Lambda^2\over 8\pi^2}-{\Lambda T\over 2\pi^2}+{T^2\over 12}\nn
&\equiv &I_{12}\Lambda^2+I_{11}\Lambda T+I_{10}T^2\nn
I_2 &=& -{1\over 16\pi^2}\ln{\Lambda\over T}+{1\over 16\pi^2}(1+\ln (2\pi )
-\gamma_E)\nn
&\equiv &I_{2ln}\ln{\Lambda\over T}+I_{20}.
\end{eqnarray}

2.
\begin{eqnarray}
I_3(m) &=& {1\over 2\pi^2}\Lambda m^2-{1\over 6\pi}m^3\nn
&\equiv & 2J_1\Lambda m^2+2J_0m^3.
\end{eqnarray}

3.
\begin{eqnarray}
H_4(m_1,m_2,m_3) &=& K_0+K_1{m_1^2+m_2^2+m_3^2\over 3}+...\nn
K_0 &=& 0.0001041333 \Lambda^2-0.0029850437\Lambda T+{5\over 32\pi^2}T^2
\ln{\Lambda\over T}\nn
&& -0.0152887686 T^2\nn
&\equiv &K_{02}\Lambda^2+K_{01}\Lambda T+K_{0ln}\ln{\Lambda\over T}
+K_{00}T^2\nn
K_1 &=& -{3\over 128\pi^4}(\ln{\Lambda\over T})^2
+0.001087871\ln{\Lambda\over T}+{\rm const.}\nn
&\equiv&K_{10}(\ln{\Lambda\over T})^2+K_{1log}\ln{\Lambda\over T}+{\rm const.}.
\end{eqnarray}

4.
\begin{eqnarray}
H_3(m_1,m_2,m_3)&=&{1\over 32\pi^2}\ln{\Lambda^2\over (m_1+m_2+m_3)^2}
+L_0,\nn
L_0&=&6.70322\times 10^{-3}.
\end{eqnarray}

5.
\begin{eqnarray}
H_{43} &=& \int_{P1}^\prime\int_{P2}^\prime\int_{\bf p}{\displaystyle 1\over
\displaystyle (P_1^2+m_1^2)(P_2^2+m_2^2)({\bf p}^2+m_3^2)}\nn
&=& a_1\Lambda T-{1\over 16\pi^2}T^2\ln{\Lambda\over T}+a_2T^2+
{m_3T\over 32\pi^3}(\ln{\Lambda\over T}-16\pi^2I_{20})\nn
&&+a_3m_3^2-{1\over 768\pi^2}(m_1^2+m_2^2)+...,
\end{eqnarray}

\begin{equation}
a_1=0.000995014555,\quad a_2=0.0074744795,\quad a_3=0.0000659642.
\end{equation}

6.
\begin{eqnarray}
L_4(m_1,m_2) &=& \int_{P1}^\prime\int_{P2}^\prime{(P_1P_2)^2
\over P_1^2(P_1^2+m_1^2)P_2^2(P_2^2+m_2^2)}\nn
&=& L_{40}+\Biggl[{T^2\over768\pi^2}-\nn
&&
-{\Lambda T\over128\pi^4}\left(\ln{\Lambda\over T}-
(1+\ln (2\pi )-\gamma_E)\right)\Biggr](m_1^2+m_2^2).
\end{eqnarray}

7.
\begin{eqnarray}
C(m_1,m_2)&=&T^2\sum_{n\neq 0}\int_{\bf k}\int_{\bf q}
{1\over (\omega_n^2+{\bf k}^2+m_1^2)}{1\over (\omega_n^2+{\bf q}^2+m_2^2)}\nn
&&=C_0+(m_1^2+m_2^2)C_1+...,
\end{eqnarray}

\begin{equation}
C_1=-{1\over 16\pi^4}\Lambda T\left(\log{\Lambda\over T}
-1-\ln (2\pi )+\gamma_E\right)-{T^2\over 32\pi^2}.
\end{equation}

\section{Momentum dependent (bubble) contributions to 2-point functions}

In this Appendix we give a list of the diagrams generating higher
 momentum-dependent vertices of the reduced theory. Each bubble contains
two internal lines, which we use for "naming" the diagrams. 
(The abbreviations are: V -- for non-static magnetic vector potential, 
H -- for non-static Higgs, G -- for non-static Goldstone, C -- for non-static 
ghost and A -- for static electric potential). Some diagrams
are proportional also to $\PU^2$, so they actually correspond to 4-point 
functions. The 2-point approach is just a convenient way to study their
momentum dependence. All 2-point functions depend on the spatial momentum
$K=(0,{\bf k})$. The first few terms of  the gradient expansion of each 
diagram will be given up to that term which should be considered in 
the evaluation
of local mass and wave function renormalisations. These terms are subtracted
from the full expression, when one turns to the investigation of non-local 
effects (see Appendix E).

{\it Higgs 2-point function}

\begin{eqnarray}
{g^2\over 8}{\bf VG} &=& -3g^2\int_P(k^2-{(KP)^2\over P^2}){1\over P^2+m_a^2}
{1\over (P+k)^2+m_G^2}\nonumber\\
&\simeq & -g^2k^2\int_P\left[3-{p^2\over P^2}\right]
{1\over P^2+m_a^2}{1\over P^2+m_G^2}
\end{eqnarray}

\begin{eqnarray}
-{g^4\over 32}\PU^2{\bf VV} & = &-{3g^4\PU^2\over 8}\int_P
{1\over P^2+m_a^2}{1\over (P+k)^2+m_a^2}\left(2+{(P(P+k))^2
\over P^2(P+k)^2}\right)\nonumber\\
&\simeq & -{9g^4\PU^2\over 8}\int_P{1\over (P^2+m_a^2)^2},
\end{eqnarray}

\begin{eqnarray}
-{\lambda^2\over 72}\PU^2{\bf HH} & = &-{\lambda^2\PU^2\over 2}\int_P
{1\over P^2+m_H^2}{1\over (P+k)^2+m_H^2}\nonumber\\
&\simeq & -{\lambda^2\PU^2\over 2}\int_P{1\over (P^2+m_H^2)^2},
\end{eqnarray}

\begin{eqnarray}
-{\lambda^2\over 72}\PU^2{\bf GG} & = &-{\lambda^2\PU^2\over 6}\int_P
{1\over P^2+m_G^2}{1\over (P+k)^2+m_G^2}\nonumber\\
&\simeq & -{\lambda^2\PU^2\over 6}\int_P{1\over (P^2+m_G^2)^2},
\end{eqnarray}

\begin{equation}
-{g^4\over 32}\PU^2{\bf AA}=-{3g^4\PU^2\over 8}\int_{\bf p}
{1\over p^2+m_{A0}^2}{1\over (p+k)^2+m_{A0}^2}.
\end{equation} 
This last integral is not analytic in $k^2$, therefore its gradient expansion
is a dangerous step.

\vskip .3truecm
{\it Goldstone 2-point function}

\begin{eqnarray}
{g^2\over 8}{\bf VG} & = -&2g^2\int_P(k^2-{(KP)^2\over P^2}){1\over P^2+m_a^2}
{1\over (P+k)^2+m_G^2}\nonumber\\
&\simeq &-2g^2k^2\int_P\left[1-{p^2\over 3P^2}\right]
{1\over P^2+m_a^2}{1\over P^2+m_G^2},
\end{eqnarray}

\begin{eqnarray}
{g^2\over 8}{\bf VH} & = -&g^2\int_P(k^2-{(KP)^2\over P^2}){1\over P^2+m_a^2}
{1\over (P+k)^2+m_H^2}\nonumber\\
&\simeq &-g^2k^2\int_P\left[1-{p^2\over 3P^2}\right]
{1\over P^2+m_a^2}{1\over P^2+m_H^2},
\end{eqnarray}

\begin{eqnarray}
-{\lambda^2\over 72}\PU^2{\bf HG} & = &-{\lambda^2\PU^2\over 9}\int_P
{1\over P^2+m_H^2}{1\over (P+k)^2+m_G^2}\nonumber\\
&\simeq &-{\lambda^2\PU^2\over 9}\int_P{1\over P^2+m_H^2}{1\over P^2+m_G^2}.
\end{eqnarray}
\vskip .3truecm
{\it Magnetic vector 2-point function}

Here we give the contributions to the polarisation tensor ${\cal N}_{ij}$. 
When the static $A_i$ legs are contracted, its contribution to the effective 
Higgs potential is of the form (see Appendix E):
\begin{equation}
\int{{\cal N}_{ij}(p)\Pi_{ij}\over p^2}-
m_{a}^2\int{({\cal N}_{ij}(p)-{\cal N}_{ij}(0))\Pi_{ij}\over p^4},
\label{eq:genform}
\end{equation}
where $\Pi_{ij}=\delta_{ij}-p_ip_j/p^2$. The following bubbles contribute:
\begin{eqnarray}
{g^2\over2} \V\V &=& -4g^2\int{1\over(K^2+m_a^2)(Q^2+m_a^2)}
\biggl[2\Pi_{ij}\left(p^2-{(pK)^2\over K^2}\right)\nonumber\\
&&+K\Pi_iK\Pi_j\biggl(3+2{pK\over K^2}+{(Kp)(Qp)\over K^2Q^2}\biggr)
\biggr]\nonumber\\
&\simeq & -4g^2\Pi_{ij}\int{k^2\over(K^2+m_a^2)^2}-4g^2p^2\Pi_{ij}\biggl[
2\int{1\over(K^2+m_a^2)^2}\nonumber\\
&&-{2\over3}\int{k^2\over K^2(K^2+m_a^2)^2}
-\int{k^2\over(K^2+m_a^2)^3}\nonumber\\
&& +{4\over5}\int{k^4\over(K^2+m_a^2)^4}
-{4\over15}\int{k^4\over K^2(K^2+m_a^2)^3}\nonumber\\
&&-{1\over15}\int{k^4\over K^4(K^2+m_a^2)^2}\biggr]\nonumber\\
{g^2\over8}\G\H &=& -g^2\int{K\Pi_iK\Pi_j\over(K^2+m_H^2)(Q^2+m_G^2)}\nonumber\\
&\simeq &-{g^2\over3}\Pi_{ij}\int{q^2\over(Q^2+m_H^2)(Q^2+m_G^2)}\nonumber\\
&&+{g^2\over3}p^2\Pi_{ij}\biggl[\int{q^2\over(Q^2+m_H^2)(Q^2+m_G^2)^2}\left(
1-{4\over5}{q^2\over Q^2+m_G^2}\right)\biggr]\nonumber\\
{g^2\over8}\G\G &=& -g^2\int{K\Pi_iK\Pi_j\over(K^2+m_G^2)(Q^2+m_G^2)}\nonumber\\
&\simeq &-{g^2\over3}\Pi_{ij}\int{k^2\over(K^2+m_G^2)^2}\nonumber\\
&&+{g^2\over3}p^2\Pi_{ij}\biggl[\int{k^2\over(K^2+m_G^2)^3}\left(
1-{4\over5}{k^2\over K^2+m_G^2}\right)\biggr]\nonumber\\
-{g^4\over32}\PU^2\,\V\H &=& -{g^4\over4}\PU^2
\int{1\over(K^2+m_a^2)(Q^2+m_H^2)}\biggl[\Pi_{ij}-{K\Pi_iK\Pi_j\over K^2}
\biggr]\nonumber\\
&\simeq &-{g^4\over4}\PU^2\Pi_{ij}\int{1\over(K^2+m_H^2)(K^2+m_a^2)}\left(1
-{1\over3}{k^2\over K^2}\right)\nonumber\\
{g^2\over2}\C\C &=& 2g^2\int{K\Pi_iK\Pi_j\over K^2 Q^2}\nonumber\\
&\simeq & {2g^2\over3}\Pi_{ij}\int{k^2\over K^4}-{2g^2\over3}p^2\Pi_{ij}
\int{k^2\over K^6}\left(1-{4\over5}{k^2\over K^2}\right)\nonumber\\
{g^2\over2}\A\A &\simeq& -{4\over3}g^2\int{k^2\over (k^2+m_{A0}^2)^2}
\left[1-{p^2\over k^2+m_{A0}^2}+{4\over5}{p^2k^2\over 
(k^2+m_{A0}^2)^2}\right].
\end{eqnarray}
($K\Pi_i\equiv k_j(\delta_{ij}-p_ip_j/p^2)).$  
\vskip .3truecm
{\it Ghost 2-point function}
\begin{equation}
{g^2\over2}\C\V =2g^2\int_K\left(p^2-{(PK)^2\over K^2}\right){1\over 
K^2+m_a^2}{1\over (P+K)^2}
\end{equation}
\section{Couplings in terms of $T=0$ physical quantities}

At tree level the renormalised couplings and the vacuum expectation value of the
Higgs-field ($v$) determine the masses of the field quanta:
\begin{eqnarray}
&
m_H^2=m^2+{\lambda\over2}v^2,\nonumber\\
&
m_W^2={g^2\over 4}v^2.
\label{masst0}
\end{eqnarray}
Physical masses are approximated by finding the roots at ${\bf p}^2=-m^2$ 
of the corresponding 2-point functions. For achieving the accuracy 
${\cal  O}(g^4,..)$ we need also the vacuum expectation value $v$ corrected
by 1-loop fluctuations. For this we calculate the minimum of the $T=0$
renormalised effective Higgs potential to 1-loop from the condition:
\begin{eqnarray}
{dV_{eff}(v)\over dv^2} &=& {1\over2}m^2(1-2\lambda I_{20})+{1\over12}\lambda
v^2(1-{27g^4\over 4\lambda}I_{20}-4\lambda I_{20})\nn
&& +{1\over 64\pi^2}\left({9g^2\over2}m_W^2\ln{m_W^2\over T^2}+\lambda 
m_G^2\ln {m_G^2\over T^2}+\lambda m_H^2\ln{m_H^2\over T^2}\right)\nn
&& =0.
\end{eqnarray}
For the accuracy requested it is sufficient to use in the 1-loop part the 
tree level expressions (\ref{masst0}). Then the iterative solution is easily 
found:
\begin{eqnarray}
v^2 &=& -{6m^2\over\lambda}\bigl(1+(2\lambda +{27g^4\over 4\lambda})I_{20}\nn
&&-{1\over 64\pi^2}[{27g^4\over 2\lambda}\ln({3g^2\over 2\lambda}(-{m^2\over
T^2}))+4\lambda \ln (-{2m^2\over T^2})]\bigr).
\end{eqnarray}

For the Higgs and W-masses one makes use of their 2-point functions.
Here again we use $\mu =T$ and neglect the running of the couplings to
the scale where they are usually measured (say at $m_W$). (This should
be used only in the tree level mass expressions, if necessary.)

The 4-d integrals were evaluated with cut-off regularisation.
We have reduced all the 1-loop integrals to 1-variable integrals with help
of Feynman parametrisation. 
For a set of numerical values of $g,\lambda$ one finds $M_H^2$ and $M_W^2$,
the 1-loop accurate masses in proportion to the renormalised mass parameter 
$-m^2$ to be still functions of the logarithms of $m_H^2/T^2$ and of 
$m_W^2/T^2$. In the arguments of the logarithms the tree level mass 
expressions can be used. This leads to the unique dependence on $-m^2/T^2$. 
The numerical value of the latter for fixed values of $g,\lambda$ 
is extracted from the degeneracy condition imposed on the minima of 
the effective potential (see
section 6). Having the value of $\ln (-m^2/T^2)$, the integrations over
the Feynman parameter can be done.  

Below we give the expressions of the 1-loop accurate Higgs and W-masses 
before the 1-dimensional integrals are performed.

The 1-loop accurate expression of the Higgs-mass is
\begin{equation}
M_H^2=m_H^2+\Sigma (-m_H^2).
\end{equation}
The self-energy of the Higgs field is given by
\begin{eqnarray}  
\Sigma (-m_H^2) &=& -{3g^2m_W^2\over 64\pi^2}(9-3h^2+{1\over 2}h^4)\ln{T^2
\over m_W^2}
-{\lambda m_H^2\over 8\pi^2}\ln{T^2\over m_H^2}\nn
&& 
+{\lambda m_H^2\over 32\pi^2}(\pi\sqrt{3}-{89\over30})+
{3g^2m_W^2\over 32\pi^2}\left(3-{7h^2\over 12}+{h^4\over 8}-Q_{H}(h^2)\right)\nn
&& +\Sigma_{CT}(-m_H^2),
\end{eqnarray}
where $h\equiv m_H/m_W$ and the one-variable integral $Q_{H}$ is defined as
\begin{eqnarray}
Q_{H}(h^2) &=& \int_0^1dx\left[1+h^2({1\over 2}-12x+12x^2)
+h^4x(-{1\over2}+{21\over2}x-20x^2+10x^3)\right]\nn
&&\times\ln (1-h^2x(1-x)).
\end{eqnarray}
The contribution coming from the 1-loop counterterms is
\begin{equation}
\Sigma_{CT}={m_H^2\over 16\pi^2}(1+\ln (2\pi )-\gamma_E)({9g^2\over 2}-
4\lambda-{81g^4\over 8\lambda})-{g^2m_H^2\over 32\pi^2}.
\end{equation}

We give the W-boson polarisation function at $p^2=-m_W^2$ in an even 
simpler form:
\begin{eqnarray}
\Pi_W(-m_W^2) &=& {g^2m_W^2\over 96\pi^2}\left({59+3h^2\over 2}\ln 
{T^2\over m_W^2}+21.188-{3h^2\over4}\right)\nn
&& -{g^2m_W^2\over 32\pi^2}\int_0^1dx[-1+(h^2-2)x+x^2]\ln [(1-x)^2+h^2x]\nn
&&+{\lambda m_W^2\over 24\pi^2}\left(1-{1\over2}
\ln{T^2\over m_H^2}\right).
\end{eqnarray}

\section{Mixed 2-loop contribution to the potential energy density of the
effective theory}

{\it 'Figure 8' diagrams and thermal counterterm:}
\begin{eqnarray}
&{g^2\over2}\A\V &=
6g^2\int{1\over k^2+m_{A0}^2}\int{1\over Q^2+m_a^2}\nn
&&+3g^2\int{1\over k^2+m_{A0}^2}\int{Q_0^2\over Q^2(Q^2+m_a^2)},\nn
&{g^2\over8}(\A\G+\A\H)& ={3g^2\over8}\int{1\over k^2+m_{A0}^2}\left(
3\int{1\over Q^2+m_G^2}+\int{1\over Q^2+m_H^2}\right),\nn
&-{1\over 2} m_D^2\A &= -{3\over2} m_D^2\int{1\over k^2+m_{A0}^2}.
\end{eqnarray}    
Here the integration signs refer either to 3-d momentum space integration over
$\bf k$, or to finite T sum-integrals over Q.
\vskip .3truecm
{\it Mixed setting-sun diagrams:}

\begin{eqnarray}
&{g^2\over2}\A\V\V &=-6g^2\int{1\over(p^2+m_{A0}^2)(Q^2+m_a^2)(K^2+m_a^2)}\nn
&&
\left(2p^2-2{(pk)^2\over K^2}+2Q_0^2+Q_0^2{(KQ)^2\over K^2Q^2}\right),\nn
&{g^2\over8}\A\H\G &=-{3\over2}g^2\int{Q_0^2\over(p^2+m_{A0}^2)(Q^2+m_H^2)
(K^2+m_G^2)},\nn
&{g^2\over8}\A\G\G &=-{3\over2}g^2\int{Q_0^2\over(p^2+m_{A0}^2)(Q^2+m_G^2)
(K^2+m_G^2)},\nn
&-{g^2\over8}\P^2\A\V\H &=-{3\over8}g^4\P^2\int{1\over(p^2+m_{A0}^2)
(Q^2+m_a^2)(K^2+m_H^2)}\left(1-{Q_0^2\over Q^2}\right),\nn
&{g^2\over2}\A\C\C &= 3g^2\int{Q_0^2\over(p^2+m_{A0}^2)Q^2K^2}.
\end{eqnarray}
In these integrals the static momentum $\bf p$ and the non-static momenta
$Q$ and $K$ are  related via ${\bf p}+K+Q=0$.

\section{Dependence of Feynman-integrals involving a static propagator on 
its mass}

We analyze in Sections 4 and 5 integrals which are of the general form
\begin{equation}
\int_{\bf p}{I(p^2)\over p^2+m^2},
\label{E1}
\end{equation}
where $I(p^2)$ is the result of non-static sum-integrals. This function 
has a well-defined Taylor-series in ${\bf p}^2$, therefore one can form the 
following set of functions:
\begin{equation}
I_s(p^2) =I(p^2)-\sum\limits_{i=0}^{s-1}{p^{2i}\over i!} I^{(i)}(0),
\qquad\qquad I_0\equiv I,
\label{E2}
\end{equation}
with $I^{(i)}$ denoting the i-th derivative with respect to ${\bf p}^2$.
The dependence of $I({\bf p}^2)$ on the non-static mass-squares is also 
analytic, the coefficients of the corresponding series expansions are IR-finite.

Below we show that the part of the above integral depending non-analytically 
on $m^2$ can be constructed explicitly. Namely, we prove, that
\begin{equation}
\int{I(p^2)\over p^2+m^2} = -{m\over4\pi}I(p^2=-m^2)
+\sum\limits_{s=0}^\infty(-m^2)^s\int{I_s(p^2)\over p^{2(s+1)}}.
\label{E3}
\end{equation}
 The first term on the right hand side is the explicit non-analytic piece, 
which can be interpreted as a factorised product of a static and of a 
non-static integral. The second term is IR-convergent.

The proof is based on the following identity:
\begin{equation}
{I_s(p^2)\over p^{2(s+1)}(p^2+m^2)}={I_{s+1}(p^2)\over p^{2(s+2)}}+
{I^{(s)}\over s!}{1\over p^2(p^2+m^2)}
-m^2{I_{s+1}(p^2)\over p^{2(s+2)}(p^2+m^2)}
\label{E4}
\end{equation}
This identity is applied repeatedly starting with the second term on the 
right hand side of the simple identity:
\begin{equation}
{I(p^2)\over p^2+m^2}={I(p^2)\over p^2}-m^2{I(p^2)\over p^2(p^2+m^2)}.
\label{E5}
\end{equation}
Eventually, one arrives at the following representation of the integrand of 
(\ref{E1}):
\begin{equation}
{I(p^2)\over p^2+m^2}=\sum\limits_{s=0}^\infty(-m^2)^s{I_s(p^2)
\over p^{2(s+1)}}\,\,-m^2\sum\limits_{s=0}^\infty(-m^2)^s{I^{(s)}(0)
\over s!}{1\over p^2(p^2+m^2)}.
\label{E6}
\end{equation}
The second term on the right hand side can be evaluated explicitly, when this 
equality is integrated with respect to ${\bf p}$.
Also the $s$-sum in it can be done, leading to (\ref{E3}).

In our applications the integral we are interested in is a "mixed" 
static--non-static "setting sun". From its two vertices one has at least a 
factor $g^2$, therefore in an ${\cal O}(g^4)$ accurate calculation one is 
satisfied with keeping the first few terms of the expansion (\ref{E6}) 
($m^2$ is of the order of $g^2$):
\begin{equation}
\int_{\bf p}{I(p^2)\over p^2+m^2}=-{m\over 4\pi}I(0)+\int_{\bf p}{I(p^2)
\over p^2}-m^2\int_{\bf p}{I_1(p^2)\over p^4}+{\cal O}(g^5).
\label{E7}
\end{equation}
On the right hand side in the second term it is sufficient for the present 
accuracy to use the expansion of
$I(p^2)$ to linear order with respect to the non-static mass squares.
In the third integral for the same purpose one simply sets these masses equal 
to zero.
\vskip .5truecm
\section*{ Acknowledgements}

The authors are grateful to Z. Fodor, M. Laine, G. Mack, I. Montvay and
J. Polonyi for valuable remarks, helpful criticism and encouragement.

{\bf Tables}

{\bf Table 1}:Critical quantities from the non-polynomial, nonlocal 
approximation (NNA)
\vskip .25truecm
{\bf Table 2}:{Percentual variation of physical quantities relative 
to the NNA approximation. The symbols $\delta_P$ and $\delta_N$ refer to the
$(LPA-NNA)/LPA$ and $(LNA-NNA)/LNA$ differences, respectively.
\vskip .2truecm
{\bf Figure Captions}

{\bf Figure 1}:$T_c/M_H$ ratio as a function of $M_H/M_W$ in NNA
\vskip .25truecm
{\bf Figure 2}:$\Phi_c/T_c$ ratio as a function of $M_H/M_W$ in NNA
\vskip .25truecm
{\bf Figure 3}:$L_c/T_c^4$ ratio as a function of $M_H/M_W$ in NNA
\vskip .25truecm
{\bf Figure 4}:The surface tension ($\sigma_c$) in $({\rm GeV})^3$ units
as a function of $M_H/M_W$ in various approximations. {\bf a}: NNA,
{\bf b}: LPA, {\bf c}: LNA

\newpage
\begin{table}
\input{tab1.tex}

\caption{Critical quantities from the non-polynomial, nonlocal approximation
(NNA)}
\end{table}

\begin{table}
\input{tab2.tex}

\caption{Percentual variation of physical quantities relative to the
NNA approximation. The symbols $\delta_P$ and $\delta_N$ refer to the 
$(LPA-NNA)/LPA$ and $(LNA-NNA)/LNA$ differences, respectively.}
\end{table}

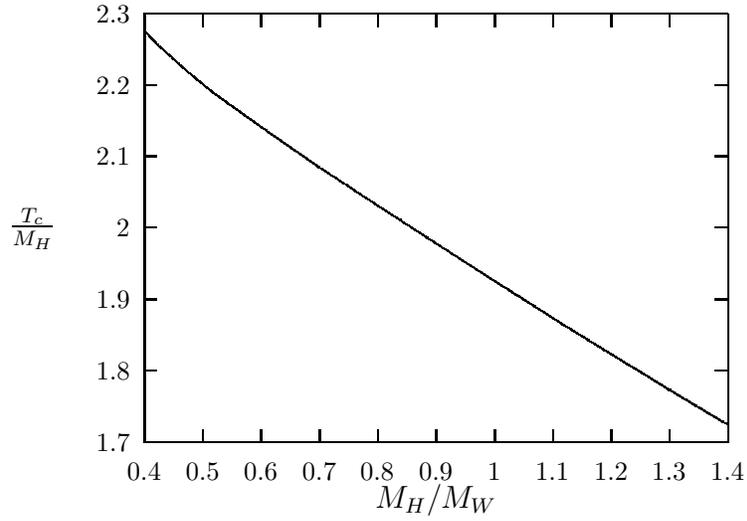
\begin{figure}[h]
\begin{center}
\input{gr1.tex}
\end{center}
\caption{$T_c/M_H$ ratio as a function of $M_H/M_W$ in NNA}
\end{figure} 

\begin{figure}[h]
\begin{center}
\input{gr2.tex}
\end{center}
\caption{$\Phi_c/T_c$ ratio as a function of $M_H/M_W$ in NNA}
\end{figure}
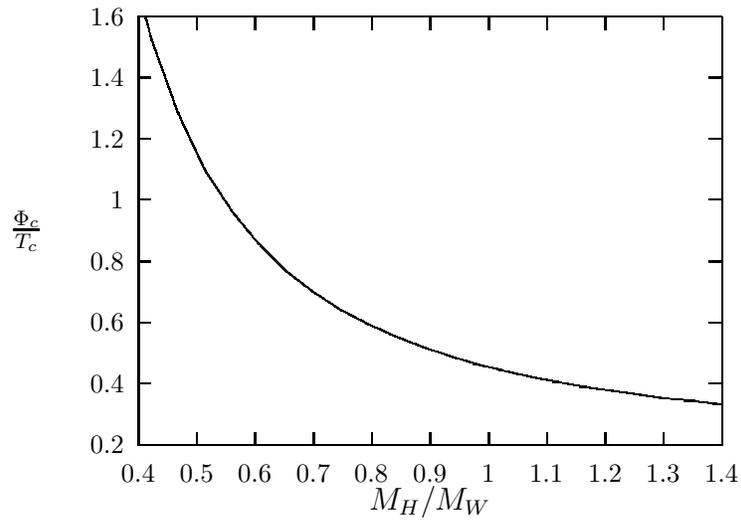

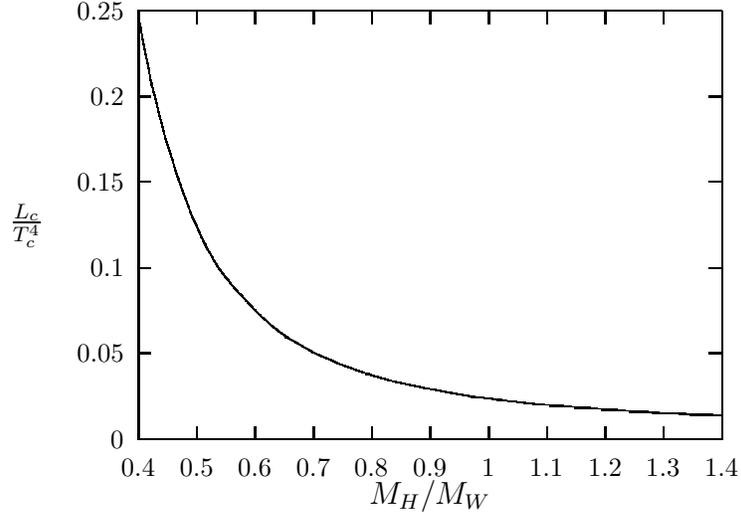
\begin{figure}[h]
\begin{center}
\input{gr4.tex}
\end{center}
\caption{$L_c/T_c^4$ ratio as a function of $M_H/M_W$ in NNA}
\end{figure}
  
\begin{figure}[h]
\begin{center}
\input{gr5.tex}
\end{center}
\caption{The surface tension ($\sigma_c$) in $({\rm GeV})^3$ units 
as a function of $M_H/M_W$ in various approximations. {\bf a}: NNA,
{\bf b}: LPA, {\bf c}: LNA}
\end{figure}
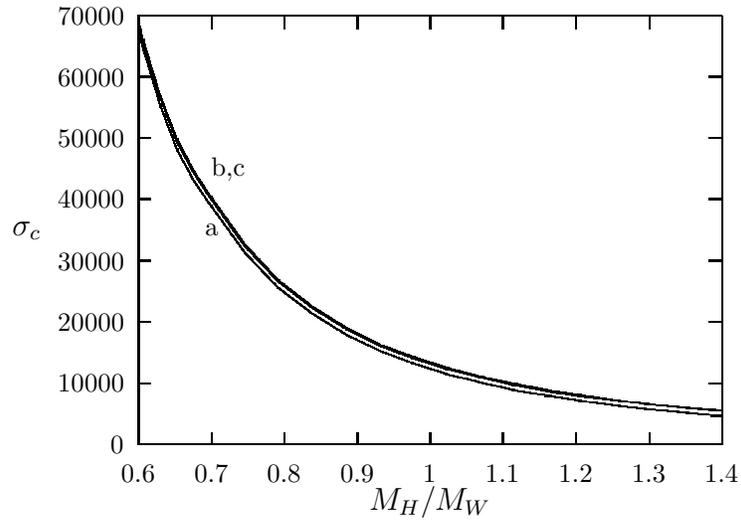

\end{document}

%% file: tab1.tex
\begin{center}
\begin{tabular}{|c|c|c|c|c|c|}
\hline\hline
$m_H^*$ & $M_H$ & $T_c/M_H$ & $\Phi_c/T_c$ & $\sigma_c/T_c^3$ 
& $L_c/T_c^4$ \\ \hline

30 &   26.72 & 2.35 & 2.336 & 0.1325 & 0.440 \\ \hline
35 &   31.85 & 2.28 & 1.719 & 0.0634 & 0.257 \\ \hline
40 &   36.96 & 2.23 & 1.333 & 0.0350 & 0.163 \\ \hline
50 &   47.16 & 2.15 & 0.906 & 0.0146 & 0.081 \\ \hline
60 &   57.28 & 2.08 & 0.687 & 0.0080 & 0.049 \\ \hline
70 &   67.30 & 2.01 & 0.558 & 0.0051 & 0.034 \\ \hline
80 &   77.23 & 1.95 & 0.476 & 0.0036 & 0.026 \\ \hline
90 &   87.07 & 1.88 & 0.419 & 0.0028 & 0.020 \\ \hline
100 &  96.81 & 1.82 & 0.378 & 0.0023 & 0.017 \\ \hline
110 & 106.45 & 1.76 & 0.348 & 0.0019 & 0.015 \\ \hline
120 & 115.99 & 1.71 & 0.325 & 0.0017 & 0.013 \\ \hline

\end{tabular}
\end{center}

%% file: tab2.tex
\begin{center}
\begin{tabular}{|c||c|c||c|c||c|c||c|c|}
\hline\hline
$m_H^*$ & $\delta_P T_c$ & $\delta_N T_c$
 & $\delta_P \Phi_c$ & $\delta_N \Phi_c$
 & $\delta_P \sigma_c$ & $\delta_N \sigma_c$
 & $\delta_P L_c$ & $\delta_N L_c$ \\ \hline

30 &   0.84 &  0.82 & -2.25 & -1.81 & -6.19 & -5.12 & -6.29 & -5.48  \\ \hline
35 &   0.88 &  0.85 & -1.89 & -1.46 & -4.50 & -3.71 & -5.69 & -4.87  \\ \hline
40 &   0.87 &  0.84 & -1.38 & -0.99 & -3.15 & -2.21 & -4.73 & -3.98  \\ \hline
50 &   0.79 &  0.76 & -0.49 & -0.10 & -0.95 & -0.17 & -3.10 & -2.35  \\ \hline
60 &   0.69 &  0.65 &  0.22 &  0.63 &  0.91 &  1.81 & -2.04 & -1.26  \\ \hline
70 &   0.57 &  0.53 &  0.96 &  1.41 &  2.47 &  3.37 & -1.22 & -0.38  \\ \hline
80 &   0.45 &  0.42 &  1.73 &  2.13 &  4.74 &  6.38 & -0.64 &  0.09  \\ \hline
90 &   0.33 &  0.30 &  2.61 &  3.05 &  6.99 &  8.55 & -0.11 &  0.66  \\ \hline
100 &  0.21 &  0.18 &  3 58 &  4.04 &  9.37 & 10.94 &  0.34 &  1.13  \\ \hline
110 &  0.09 &  0.06 &  4.50 &  5.08 & 12.93 & 13.59 &  0.45 &  1.44  \\ \hline
120 & -0.02 & -0.06 &  5.59 &  6.11 & 15.56 & 16.99 &  0.62 &  1.45  \\ \hline

\end{tabular}
\end{center}

%% file: gr1.tex
\setlength{\unitlength}{0.240900pt}
\ifx\plotpoint\undefined\newsavebox{\plotpoint}\fi
\sbox{\plotpoint}{\rule[-0.200pt]{0.400pt}{0.400pt}}%
\begin{picture}(1200,809)(0,0)
\font\gnuplot=cmr10 at 10pt
\gnuplot
\sbox{\plotpoint}{\rule[-0.200pt]{0.400pt}{0.400pt}}%
\put(220.0,113.0){\rule[-0.200pt]{4.818pt}{0.400pt}}
\put(198,113){\makebox(0,0)[r]{1.7}}
\put(1116.0,113.0){\rule[-0.200pt]{4.818pt}{0.400pt}}
\put(220.0,225.0){\rule[-0.200pt]{4.818pt}{0.400pt}}
\put(198,225){\makebox(0,0)[r]{1.8}}
\put(1116.0,225.0){\rule[-0.200pt]{4.818pt}{0.400pt}}
\put(220.0,337.0){\rule[-0.200pt]{4.818pt}{0.400pt}}
\put(198,337){\makebox(0,0)[r]{1.9}}
\put(1116.0,337.0){\rule[-0.200pt]{4.818pt}{0.400pt}}
\put(220.0,450.0){\rule[-0.200pt]{4.818pt}{0.400pt}}
\put(198,450){\makebox(0,0)[r]{2}}
\put(1116.0,450.0){\rule[-0.200pt]{4.818pt}{0.400pt}}
\put(220.0,562.0){\rule[-0.200pt]{4.818pt}{0.400pt}}
\put(198,562){\makebox(0,0)[r]{2.1}}
\put(1116.0,562.0){\rule[-0.200pt]{4.818pt}{0.400pt}}
\put(220.0,674.0){\rule[-0.200pt]{4.818pt}{0.400pt}}
\put(198,674){\makebox(0,0)[r]{2.2}}
\put(1116.0,674.0){\rule[-0.200pt]{4.818pt}{0.400pt}}
\put(220.0,786.0){\rule[-0.200pt]{4.818pt}{0.400pt}}
\put(198,786){\makebox(0,0)[r]{2.3}}
\put(1116.0,786.0){\rule[-0.200pt]{4.818pt}{0.400pt}}
\put(220.0,113.0){\rule[-0.200pt]{0.400pt}{4.818pt}}
\put(220,68){\makebox(0,0){0.4}}
\put(220.0,766.0){\rule[-0.200pt]{0.400pt}{4.818pt}}
\put(312.0,113.0){\rule[-0.200pt]{0.400pt}{4.818pt}}
\put(312,68){\makebox(0,0){0.5}}
\put(312.0,766.0){\rule[-0.200pt]{0.400pt}{4.818pt}}
\put(403.0,113.0){\rule[-0.200pt]{0.400pt}{4.818pt}}
\put(403,68){\makebox(0,0){0.6}}
\put(403.0,766.0){\rule[-0.200pt]{0.400pt}{4.818pt}}
\put(495.0,113.0){\rule[-0.200pt]{0.400pt}{4.818pt}}
\put(495,68){\makebox(0,0){0.7}}
\put(495.0,766.0){\rule[-0.200pt]{0.400pt}{4.818pt}}
\put(586.0,113.0){\rule[-0.200pt]{0.400pt}{4.818pt}}
\put(586,68){\makebox(0,0){0.8}}
\put(586.0,766.0){\rule[-0.200pt]{0.400pt}{4.818pt}}
\put(678.0,113.0){\rule[-0.200pt]{0.400pt}{4.818pt}}
\put(678,68){\makebox(0,0){0.9}}
\put(678.0,766.0){\rule[-0.200pt]{0.400pt}{4.818pt}}
\put(770.0,113.0){\rule[-0.200pt]{0.400pt}{4.818pt}}
\put(770,68){\makebox(0,0){1}}
\put(770.0,766.0){\rule[-0.200pt]{0.400pt}{4.818pt}}
\put(861.0,113.0){\rule[-0.200pt]{0.400pt}{4.818pt}}
\put(861,68){\makebox(0,0){1.1}}
\put(861.0,766.0){\rule[-0.200pt]{0.400pt}{4.818pt}}
\put(953.0,113.0){\rule[-0.200pt]{0.400pt}{4.818pt}}
\put(953,68){\makebox(0,0){1.2}}
\put(953.0,766.0){\rule[-0.200pt]{0.400pt}{4.818pt}}
\put(1044.0,113.0){\rule[-0.200pt]{0.400pt}{4.818pt}}
\put(1044,68){\makebox(0,0){1.3}}
\put(1044.0,766.0){\rule[-0.200pt]{0.400pt}{4.818pt}}
\put(1136.0,113.0){\rule[-0.200pt]{0.400pt}{4.818pt}}
\put(1136,68){\makebox(0,0){1.4}}
\put(1136.0,766.0){\rule[-0.200pt]{0.400pt}{4.818pt}}
\put(220.0,113.0){\rule[-0.200pt]{220.664pt}{0.400pt}}
\put(1136.0,113.0){\rule[-0.200pt]{0.400pt}{162.126pt}}
\put(220.0,786.0){\rule[-0.200pt]{220.664pt}{0.400pt}}
\put(45,449){\makebox(0,0){${T_c\over M_H}$}}
\put(678,23){\makebox(0,0){$M_H/M_W$}}
\put(220.0,113.0){\rule[-0.200pt]{0.400pt}{162.126pt}}
\multiput(220.58,758.77)(0.496,-0.544){41}{\rule{0.120pt}{0.536pt}}
\multiput(219.17,759.89)(22.000,-22.887){2}{\rule{0.400pt}{0.268pt}}
\multiput(242.00,735.92)(0.552,-0.498){73}{\rule{0.542pt}{0.120pt}}
\multiput(242.00,736.17)(40.875,-38.000){2}{\rule{0.271pt}{0.400pt}}
\multiput(284.00,697.92)(0.600,-0.498){67}{\rule{0.580pt}{0.120pt}}
\multiput(284.00,698.17)(40.796,-35.000){2}{\rule{0.290pt}{0.400pt}}
\multiput(326.00,662.92)(0.695,-0.497){59}{\rule{0.655pt}{0.120pt}}
\multiput(326.00,663.17)(41.641,-31.000){2}{\rule{0.327pt}{0.400pt}}
\multiput(369.00,631.92)(0.701,-0.497){57}{\rule{0.660pt}{0.120pt}}
\multiput(369.00,632.17)(40.630,-30.000){2}{\rule{0.330pt}{0.400pt}}
\multiput(411.00,601.92)(0.726,-0.497){55}{\rule{0.679pt}{0.120pt}}
\multiput(411.00,602.17)(40.590,-29.000){2}{\rule{0.340pt}{0.400pt}}
\multiput(453.00,572.92)(0.726,-0.497){55}{\rule{0.679pt}{0.120pt}}
\multiput(453.00,573.17)(40.590,-29.000){2}{\rule{0.340pt}{0.400pt}}
\multiput(495.00,543.92)(0.770,-0.497){53}{\rule{0.714pt}{0.120pt}}
\multiput(495.00,544.17)(41.517,-28.000){2}{\rule{0.357pt}{0.400pt}}
\multiput(538.00,515.92)(0.752,-0.497){53}{\rule{0.700pt}{0.120pt}}
\multiput(538.00,516.17)(40.547,-28.000){2}{\rule{0.350pt}{0.400pt}}
\multiput(580.00,487.92)(0.780,-0.497){51}{\rule{0.722pt}{0.120pt}}
\multiput(580.00,488.17)(40.501,-27.000){2}{\rule{0.361pt}{0.400pt}}
\multiput(622.00,460.92)(0.770,-0.497){53}{\rule{0.714pt}{0.120pt}}
\multiput(622.00,461.17)(41.517,-28.000){2}{\rule{0.357pt}{0.400pt}}
\multiput(665.00,432.92)(0.780,-0.497){51}{\rule{0.722pt}{0.120pt}}
\multiput(665.00,433.17)(40.501,-27.000){2}{\rule{0.361pt}{0.400pt}}
\multiput(707.00,405.92)(0.780,-0.497){51}{\rule{0.722pt}{0.120pt}}
\multiput(707.00,406.17)(40.501,-27.000){2}{\rule{0.361pt}{0.400pt}}
\multiput(749.00,378.92)(0.780,-0.497){51}{\rule{0.722pt}{0.120pt}}
\multiput(749.00,379.17)(40.501,-27.000){2}{\rule{0.361pt}{0.400pt}}
\multiput(791.00,351.92)(0.799,-0.497){51}{\rule{0.737pt}{0.120pt}}
\multiput(791.00,352.17)(41.470,-27.000){2}{\rule{0.369pt}{0.400pt}}
\multiput(834.00,324.92)(0.780,-0.497){51}{\rule{0.722pt}{0.120pt}}
\multiput(834.00,325.17)(40.501,-27.000){2}{\rule{0.361pt}{0.400pt}}
\multiput(876.00,297.92)(0.810,-0.497){49}{\rule{0.746pt}{0.120pt}}
\multiput(876.00,298.17)(40.451,-26.000){2}{\rule{0.373pt}{0.400pt}}
\multiput(918.00,271.92)(0.810,-0.497){49}{\rule{0.746pt}{0.120pt}}
\multiput(918.00,272.17)(40.451,-26.000){2}{\rule{0.373pt}{0.400pt}}
\multiput(960.00,245.92)(0.830,-0.497){49}{\rule{0.762pt}{0.120pt}}
\multiput(960.00,246.17)(41.419,-26.000){2}{\rule{0.381pt}{0.400pt}}
\multiput(1003.00,219.92)(0.810,-0.497){49}{\rule{0.746pt}{0.120pt}}
\multiput(1003.00,220.17)(40.451,-26.000){2}{\rule{0.373pt}{0.400pt}}
\multiput(1045.00,193.92)(0.843,-0.497){47}{\rule{0.772pt}{0.120pt}}
\multiput(1045.00,194.17)(40.398,-25.000){2}{\rule{0.386pt}{0.400pt}}
\multiput(1087.00,168.92)(0.843,-0.497){47}{\rule{0.772pt}{0.120pt}}
\multiput(1087.00,169.17)(40.398,-25.000){2}{\rule{0.386pt}{0.400pt}}
\multiput(1129.00,143.94)(0.920,-0.468){5}{\rule{0.800pt}{0.113pt}}
\multiput(1129.00,144.17)(5.340,-4.000){2}{\rule{0.400pt}{0.400pt}}
\end{picture}

%% file: gr2.tex
\setlength{\unitlength}{0.240900pt}
\ifx\plotpoint\undefined\newsavebox{\plotpoint}\fi
\begin{picture}(1200,809)(0,0)
\font\gnuplot=cmr10 at 10pt
\gnuplot
\sbox{\plotpoint}{\rule[-0.200pt]{0.400pt}{0.400pt}}%
\put(220.0,113.0){\rule[-0.200pt]{4.818pt}{0.400pt}}
\put(198,113){\makebox(0,0)[r]{0.2}}
\put(1116.0,113.0){\rule[-0.200pt]{4.818pt}{0.400pt}}
\put(220.0,209.0){\rule[-0.200pt]{4.818pt}{0.400pt}}
\put(198,209){\makebox(0,0)[r]{0.4}}
\put(1116.0,209.0){\rule[-0.200pt]{4.818pt}{0.400pt}}
\put(220.0,305.0){\rule[-0.200pt]{4.818pt}{0.400pt}}
\put(198,305){\makebox(0,0)[r]{0.6}}
\put(1116.0,305.0){\rule[-0.200pt]{4.818pt}{0.400pt}}
\put(220.0,401.0){\rule[-0.200pt]{4.818pt}{0.400pt}}
\put(198,401){\makebox(0,0)[r]{0.8}}
\put(1116.0,401.0){\rule[-0.200pt]{4.818pt}{0.400pt}}
\put(220.0,498.0){\rule[-0.200pt]{4.818pt}{0.400pt}}
\put(198,498){\makebox(0,0)[r]{1}}
\put(1116.0,498.0){\rule[-0.200pt]{4.818pt}{0.400pt}}
\put(220.0,594.0){\rule[-0.200pt]{4.818pt}{0.400pt}}
\put(198,594){\makebox(0,0)[r]{1.2}}
\put(1116.0,594.0){\rule[-0.200pt]{4.818pt}{0.400pt}}
\put(220.0,690.0){\rule[-0.200pt]{4.818pt}{0.400pt}}
\put(198,690){\makebox(0,0)[r]{1.4}}
\put(1116.0,690.0){\rule[-0.200pt]{4.818pt}{0.400pt}}
\put(220.0,786.0){\rule[-0.200pt]{4.818pt}{0.400pt}}
\put(198,786){\makebox(0,0)[r]{1.6}}
\put(1116.0,786.0){\rule[-0.200pt]{4.818pt}{0.400pt}}
\put(220.0,113.0){\rule[-0.200pt]{0.400pt}{4.818pt}}
\put(220,68){\makebox(0,0){0.4}}
\put(220.0,766.0){\rule[-0.200pt]{0.400pt}{4.818pt}}
\put(312.0,113.0){\rule[-0.200pt]{0.400pt}{4.818pt}}
\put(312,68){\makebox(0,0){0.5}}
\put(312.0,766.0){\rule[-0.200pt]{0.400pt}{4.818pt}}
\put(403.0,113.0){\rule[-0.200pt]{0.400pt}{4.818pt}}
\put(403,68){\makebox(0,0){0.6}}
\put(403.0,766.0){\rule[-0.200pt]{0.400pt}{4.818pt}}
\put(495.0,113.0){\rule[-0.200pt]{0.400pt}{4.818pt}}
\put(495,68){\makebox(0,0){0.7}}
\put(495.0,766.0){\rule[-0.200pt]{0.400pt}{4.818pt}}
\put(586.0,113.0){\rule[-0.200pt]{0.400pt}{4.818pt}}
\put(586,68){\makebox(0,0){0.8}}
\put(586.0,766.0){\rule[-0.200pt]{0.400pt}{4.818pt}}
\put(678.0,113.0){\rule[-0.200pt]{0.400pt}{4.818pt}}
\put(678,68){\makebox(0,0){0.9}}
\put(678.0,766.0){\rule[-0.200pt]{0.400pt}{4.818pt}}
\put(770.0,113.0){\rule[-0.200pt]{0.400pt}{4.818pt}}
\put(770,68){\makebox(0,0){1}}
\put(770.0,766.0){\rule[-0.200pt]{0.400pt}{4.818pt}}
\put(861.0,113.0){\rule[-0.200pt]{0.400pt}{4.818pt}}
\put(861,68){\makebox(0,0){1.1}}
\put(861.0,766.0){\rule[-0.200pt]{0.400pt}{4.818pt}}
\put(953.0,113.0){\rule[-0.200pt]{0.400pt}{4.818pt}}
\put(953,68){\makebox(0,0){1.2}}
\put(953.0,766.0){\rule[-0.200pt]{0.400pt}{4.818pt}}
\put(1044.0,113.0){\rule[-0.200pt]{0.400pt}{4.818pt}}
\put(1044,68){\makebox(0,0){1.3}}
\put(1044.0,766.0){\rule[-0.200pt]{0.400pt}{4.818pt}}
\put(1136.0,113.0){\rule[-0.200pt]{0.400pt}{4.818pt}}
\put(1136,68){\makebox(0,0){1.4}}
\put(1136.0,766.0){\rule[-0.200pt]{0.400pt}{4.818pt}}
\put(220.0,113.0){\rule[-0.200pt]{220.664pt}{0.400pt}}
\put(1136.0,113.0){\rule[-0.200pt]{0.400pt}{162.126pt}}
\put(220.0,786.0){\rule[-0.200pt]{220.664pt}{0.400pt}}
\put(45,449){\makebox(0,0){${\Phi_c\over T_c}$}}
\put(678,23){\makebox(0,0){$M_H/M_W$}}
\put(220.0,113.0){\rule[-0.200pt]{0.400pt}{162.126pt}}
\multiput(232.58,779.28)(0.491,-1.955){17}{\rule{0.118pt}{1.620pt}}
\multiput(231.17,782.64)(10.000,-34.638){2}{\rule{0.400pt}{0.810pt}}
\multiput(242.58,743.00)(0.498,-1.387){81}{\rule{0.120pt}{1.205pt}}
\multiput(241.17,745.50)(42.000,-113.499){2}{\rule{0.400pt}{0.602pt}}
\multiput(284.58,628.07)(0.498,-1.063){81}{\rule{0.120pt}{0.948pt}}
\multiput(283.17,630.03)(42.000,-87.033){2}{\rule{0.400pt}{0.474pt}}
\multiput(326.58,540.07)(0.498,-0.757){83}{\rule{0.120pt}{0.705pt}}
\multiput(325.17,541.54)(43.000,-63.537){2}{\rule{0.400pt}{0.352pt}}
\multiput(369.58,475.57)(0.498,-0.607){81}{\rule{0.120pt}{0.586pt}}
\multiput(368.17,476.78)(42.000,-49.784){2}{\rule{0.400pt}{0.293pt}}
\multiput(411.00,425.92)(0.499,-0.498){81}{\rule{0.500pt}{0.120pt}}
\multiput(411.00,426.17)(40.962,-42.000){2}{\rule{0.250pt}{0.400pt}}
\multiput(453.00,383.92)(0.657,-0.497){61}{\rule{0.625pt}{0.120pt}}
\multiput(453.00,384.17)(40.703,-32.000){2}{\rule{0.313pt}{0.400pt}}
\multiput(495.00,351.92)(0.770,-0.497){53}{\rule{0.714pt}{0.120pt}}
\multiput(495.00,352.17)(41.517,-28.000){2}{\rule{0.357pt}{0.400pt}}
\multiput(538.00,323.92)(0.960,-0.496){41}{\rule{0.864pt}{0.120pt}}
\multiput(538.00,324.17)(40.207,-22.000){2}{\rule{0.432pt}{0.400pt}}
\multiput(580.00,301.92)(1.115,-0.495){35}{\rule{0.984pt}{0.119pt}}
\multiput(580.00,302.17)(39.957,-19.000){2}{\rule{0.492pt}{0.400pt}}
\multiput(622.00,282.92)(1.279,-0.495){31}{\rule{1.112pt}{0.119pt}}
\multiput(622.00,283.17)(40.692,-17.000){2}{\rule{0.556pt}{0.400pt}}
\multiput(665.00,265.92)(1.525,-0.494){25}{\rule{1.300pt}{0.119pt}}
\multiput(665.00,266.17)(39.302,-14.000){2}{\rule{0.650pt}{0.400pt}}
\multiput(707.00,251.92)(1.646,-0.493){23}{\rule{1.392pt}{0.119pt}}
\multiput(707.00,252.17)(39.110,-13.000){2}{\rule{0.696pt}{0.400pt}}
\multiput(749.00,238.92)(2.163,-0.491){17}{\rule{1.780pt}{0.118pt}}
\multiput(749.00,239.17)(38.306,-10.000){2}{\rule{0.890pt}{0.400pt}}
\multiput(791.00,228.92)(2.215,-0.491){17}{\rule{1.820pt}{0.118pt}}
\multiput(791.00,229.17)(39.222,-10.000){2}{\rule{0.910pt}{0.400pt}}
\multiput(834.00,218.93)(2.739,-0.488){13}{\rule{2.200pt}{0.117pt}}
\multiput(834.00,219.17)(37.434,-8.000){2}{\rule{1.100pt}{0.400pt}}
\multiput(876.00,210.93)(2.739,-0.488){13}{\rule{2.200pt}{0.117pt}}
\multiput(876.00,211.17)(37.434,-8.000){2}{\rule{1.100pt}{0.400pt}}
\multiput(918.00,202.93)(3.745,-0.482){9}{\rule{2.900pt}{0.116pt}}
\multiput(918.00,203.17)(35.981,-6.000){2}{\rule{1.450pt}{0.400pt}}
\multiput(960.00,196.93)(3.836,-0.482){9}{\rule{2.967pt}{0.116pt}}
\multiput(960.00,197.17)(36.843,-6.000){2}{\rule{1.483pt}{0.400pt}}
\multiput(1003.00,190.93)(3.745,-0.482){9}{\rule{2.900pt}{0.116pt}}
\multiput(1003.00,191.17)(35.981,-6.000){2}{\rule{1.450pt}{0.400pt}}
\multiput(1045.00,184.94)(6.038,-0.468){5}{\rule{4.300pt}{0.113pt}}
\multiput(1045.00,185.17)(33.075,-4.000){2}{\rule{2.150pt}{0.400pt}}
\multiput(1087.00,180.93)(4.606,-0.477){7}{\rule{3.460pt}{0.115pt}}
\multiput(1087.00,181.17)(34.819,-5.000){2}{\rule{1.730pt}{0.400pt}}
\put(1129.0,177.0){\rule[-0.200pt]{1.686pt}{0.400pt}}
\end{picture}

%% file: gr4.tex
\setlength{\unitlength}{0.240900pt}
\ifx\plotpoint\undefined\newsavebox{\plotpoint}\fi
\begin{picture}(1200,809)(0,0)
\font\gnuplot=cmr10 at 10pt
\gnuplot
\sbox{\plotpoint}{\rule[-0.200pt]{0.400pt}{0.400pt}}%
\put(220.0,113.0){\rule[-0.200pt]{220.664pt}{0.400pt}}
\put(220.0,113.0){\rule[-0.200pt]{4.818pt}{0.400pt}}
\put(198,113){\makebox(0,0)[r]{0}}
\put(1116.0,113.0){\rule[-0.200pt]{4.818pt}{0.400pt}}
\put(220.0,248.0){\rule[-0.200pt]{4.818pt}{0.400pt}}
\put(198,248){\makebox(0,0)[r]{0.05}}
\put(1116.0,248.0){\rule[-0.200pt]{4.818pt}{0.400pt}}
\put(220.0,382.0){\rule[-0.200pt]{4.818pt}{0.400pt}}
\put(198,382){\makebox(0,0)[r]{0.1}}
\put(1116.0,382.0){\rule[-0.200pt]{4.818pt}{0.400pt}}
\put(220.0,517.0){\rule[-0.200pt]{4.818pt}{0.400pt}}
\put(198,517){\makebox(0,0)[r]{0.15}}
\put(1116.0,517.0){\rule[-0.200pt]{4.818pt}{0.400pt}}
\put(220.0,651.0){\rule[-0.200pt]{4.818pt}{0.400pt}}
\put(198,651){\makebox(0,0)[r]{0.2}}
\put(1116.0,651.0){\rule[-0.200pt]{4.818pt}{0.400pt}}
\put(220.0,786.0){\rule[-0.200pt]{4.818pt}{0.400pt}}
\put(198,786){\makebox(0,0)[r]{0.25}}
\put(1116.0,786.0){\rule[-0.200pt]{4.818pt}{0.400pt}}
\put(220.0,113.0){\rule[-0.200pt]{0.400pt}{4.818pt}}
\put(220,68){\makebox(0,0){0.4}}
\put(220.0,766.0){\rule[-0.200pt]{0.400pt}{4.818pt}}
\put(312.0,113.0){\rule[-0.200pt]{0.400pt}{4.818pt}}
\put(312,68){\makebox(0,0){0.5}}
\put(312.0,766.0){\rule[-0.200pt]{0.400pt}{4.818pt}}
\put(403.0,113.0){\rule[-0.200pt]{0.400pt}{4.818pt}}
\put(403,68){\makebox(0,0){0.6}}
\put(403.0,766.0){\rule[-0.200pt]{0.400pt}{4.818pt}}
\put(495.0,113.0){\rule[-0.200pt]{0.400pt}{4.818pt}}
\put(495,68){\makebox(0,0){0.7}}
\put(495.0,766.0){\rule[-0.200pt]{0.400pt}{4.818pt}}
\put(586.0,113.0){\rule[-0.200pt]{0.400pt}{4.818pt}}
\put(586,68){\makebox(0,0){0.8}}
\put(586.0,766.0){\rule[-0.200pt]{0.400pt}{4.818pt}}
\put(678.0,113.0){\rule[-0.200pt]{0.400pt}{4.818pt}}
\put(678,68){\makebox(0,0){0.9}}
\put(678.0,766.0){\rule[-0.200pt]{0.400pt}{4.818pt}}
\put(770.0,113.0){\rule[-0.200pt]{0.400pt}{4.818pt}}
\put(770,68){\makebox(0,0){1}}
\put(770.0,766.0){\rule[-0.200pt]{0.400pt}{4.818pt}}
\put(861.0,113.0){\rule[-0.200pt]{0.400pt}{4.818pt}}
\put(861,68){\makebox(0,0){1.1}}
\put(861.0,766.0){\rule[-0.200pt]{0.400pt}{4.818pt}}
\put(953.0,113.0){\rule[-0.200pt]{0.400pt}{4.818pt}}
\put(953,68){\makebox(0,0){1.2}}
\put(953.0,766.0){\rule[-0.200pt]{0.400pt}{4.818pt}}
\put(1044.0,113.0){\rule[-0.200pt]{0.400pt}{4.818pt}}
\put(1044,68){\makebox(0,0){1.3}}
\put(1044.0,766.0){\rule[-0.200pt]{0.400pt}{4.818pt}}
\put(1136.0,113.0){\rule[-0.200pt]{0.400pt}{4.818pt}}
\put(1136,68){\makebox(0,0){1.4}}
\put(1136.0,766.0){\rule[-0.200pt]{0.400pt}{4.818pt}}
\put(220.0,113.0){\rule[-0.200pt]{220.664pt}{0.400pt}}
\put(1136.0,113.0){\rule[-0.200pt]{0.400pt}{162.126pt}}
\put(220.0,786.0){\rule[-0.200pt]{220.664pt}{0.400pt}}
\put(45,449){\makebox(0,0){${L_c\over T_c^4}$}}
\put(678,23){\makebox(0,0){$M_H/M_W$}}
\put(220.0,113.0){\rule[-0.200pt]{0.400pt}{162.126pt}}
\put(219.67,775){\rule{0.400pt}{1.204pt}}
\multiput(219.17,777.50)(1.000,-2.500){2}{\rule{0.400pt}{0.602pt}}
\multiput(221.58,766.20)(0.496,-2.558){39}{\rule{0.119pt}{2.119pt}}
\multiput(220.17,770.60)(21.000,-101.602){2}{\rule{0.400pt}{1.060pt}}
\multiput(242.58,662.10)(0.496,-1.976){39}{\rule{0.119pt}{1.662pt}}
\multiput(241.17,665.55)(21.000,-78.551){2}{\rule{0.400pt}{0.831pt}}
\multiput(263.58,581.29)(0.496,-1.613){39}{\rule{0.119pt}{1.376pt}}
\multiput(262.17,584.14)(21.000,-64.144){2}{\rule{0.400pt}{0.688pt}}
\multiput(284.58,515.08)(0.496,-1.370){39}{\rule{0.119pt}{1.186pt}}
\multiput(283.17,517.54)(21.000,-54.539){2}{\rule{0.400pt}{0.593pt}}
\multiput(305.58,458.95)(0.496,-1.104){39}{\rule{0.119pt}{0.976pt}}
\multiput(304.17,460.97)(21.000,-43.974){2}{\rule{0.400pt}{0.488pt}}
\multiput(326.58,413.74)(0.496,-0.862){39}{\rule{0.119pt}{0.786pt}}
\multiput(325.17,415.37)(21.000,-34.369){2}{\rule{0.400pt}{0.393pt}}
\multiput(347.58,378.47)(0.496,-0.637){41}{\rule{0.120pt}{0.609pt}}
\multiput(346.17,379.74)(22.000,-26.736){2}{\rule{0.400pt}{0.305pt}}
\multiput(369.58,350.81)(0.498,-0.535){81}{\rule{0.120pt}{0.529pt}}
\multiput(368.17,351.90)(42.000,-43.903){2}{\rule{0.400pt}{0.264pt}}
\multiput(411.00,306.92)(0.551,-0.495){35}{\rule{0.542pt}{0.119pt}}
\multiput(411.00,307.17)(19.875,-19.000){2}{\rule{0.271pt}{0.400pt}}
\multiput(432.00,287.92)(0.657,-0.494){29}{\rule{0.625pt}{0.119pt}}
\multiput(432.00,288.17)(19.703,-16.000){2}{\rule{0.313pt}{0.400pt}}
\multiput(453.00,271.92)(0.879,-0.496){45}{\rule{0.800pt}{0.120pt}}
\multiput(453.00,272.17)(40.340,-24.000){2}{\rule{0.400pt}{0.400pt}}
\multiput(495.00,247.92)(1.142,-0.495){35}{\rule{1.005pt}{0.119pt}}
\multiput(495.00,248.17)(40.914,-19.000){2}{\rule{0.503pt}{0.400pt}}
\multiput(538.00,228.92)(1.421,-0.494){27}{\rule{1.220pt}{0.119pt}}
\multiput(538.00,229.17)(39.468,-15.000){2}{\rule{0.610pt}{0.400pt}}
\multiput(580.00,213.92)(1.789,-0.492){21}{\rule{1.500pt}{0.119pt}}
\multiput(580.00,214.17)(38.887,-12.000){2}{\rule{0.750pt}{0.400pt}}
\multiput(622.00,201.93)(2.475,-0.489){15}{\rule{2.011pt}{0.118pt}}
\multiput(622.00,202.17)(38.826,-9.000){2}{\rule{1.006pt}{0.400pt}}
\multiput(665.00,192.93)(2.739,-0.488){13}{\rule{2.200pt}{0.117pt}}
\multiput(665.00,193.17)(37.434,-8.000){2}{\rule{1.100pt}{0.400pt}}
\multiput(707.00,184.93)(3.162,-0.485){11}{\rule{2.500pt}{0.117pt}}
\multiput(707.00,185.17)(36.811,-7.000){2}{\rule{1.250pt}{0.400pt}}
\multiput(749.00,177.93)(4.606,-0.477){7}{\rule{3.460pt}{0.115pt}}
\multiput(749.00,178.17)(34.819,-5.000){2}{\rule{1.730pt}{0.400pt}}
\multiput(791.00,172.93)(4.718,-0.477){7}{\rule{3.540pt}{0.115pt}}
\multiput(791.00,173.17)(35.653,-5.000){2}{\rule{1.770pt}{0.400pt}}
\multiput(834.00,167.94)(6.038,-0.468){5}{\rule{4.300pt}{0.113pt}}
\multiput(834.00,168.17)(33.075,-4.000){2}{\rule{2.150pt}{0.400pt}}
\multiput(876.00,163.95)(9.169,-0.447){3}{\rule{5.700pt}{0.108pt}}
\multiput(876.00,164.17)(30.169,-3.000){2}{\rule{2.850pt}{0.400pt}}
\multiput(918.00,160.95)(9.169,-0.447){3}{\rule{5.700pt}{0.108pt}}
\multiput(918.00,161.17)(30.169,-3.000){2}{\rule{2.850pt}{0.400pt}}
\multiput(960.00,157.95)(9.393,-0.447){3}{\rule{5.833pt}{0.108pt}}
\multiput(960.00,158.17)(30.893,-3.000){2}{\rule{2.917pt}{0.400pt}}
\put(1003,154.17){\rule{8.500pt}{0.400pt}}
\multiput(1003.00,155.17)(24.358,-2.000){2}{\rule{4.250pt}{0.400pt}}
\put(1045,152.17){\rule{8.500pt}{0.400pt}}
\multiput(1045.00,153.17)(24.358,-2.000){2}{\rule{4.250pt}{0.400pt}}
\put(1087,150.17){\rule{8.500pt}{0.400pt}}
\multiput(1087.00,151.17)(24.358,-2.000){2}{\rule{4.250pt}{0.400pt}}
\put(1129.0,150.0){\rule[-0.200pt]{1.686pt}{0.400pt}}
\end{picture}

%% file: gr5.tex
\setlength{\unitlength}{0.240900pt}
\ifx\plotpoint\undefined\newsavebox{\plotpoint}\fi
\begin{picture}(1200,809)(0,0)
\font\gnuplot=cmr10 at 10pt
\gnuplot
\sbox{\plotpoint}{\rule[-0.200pt]{0.400pt}{0.400pt}}%
\put(220.0,113.0){\rule[-0.200pt]{220.664pt}{0.400pt}}
\put(220.0,113.0){\rule[-0.200pt]{4.818pt}{0.400pt}}
\put(198,113){\makebox(0,0)[r]{0}}
\put(1116.0,113.0){\rule[-0.200pt]{4.818pt}{0.400pt}}
\put(220.0,209.0){\rule[-0.200pt]{4.818pt}{0.400pt}}
\put(198,209){\makebox(0,0)[r]{10000}}
\put(1116.0,209.0){\rule[-0.200pt]{4.818pt}{0.400pt}}
\put(220.0,305.0){\rule[-0.200pt]{4.818pt}{0.400pt}}
\put(198,305){\makebox(0,0)[r]{20000}}
\put(1116.0,305.0){\rule[-0.200pt]{4.818pt}{0.400pt}}
\put(220.0,401.0){\rule[-0.200pt]{4.818pt}{0.400pt}}
\put(198,401){\makebox(0,0)[r]{30000}}
\put(1116.0,401.0){\rule[-0.200pt]{4.818pt}{0.400pt}}
\put(220.0,498.0){\rule[-0.200pt]{4.818pt}{0.400pt}}
\put(198,498){\makebox(0,0)[r]{40000}}
\put(1116.0,498.0){\rule[-0.200pt]{4.818pt}{0.400pt}}
\put(220.0,594.0){\rule[-0.200pt]{4.818pt}{0.400pt}}
\put(198,594){\makebox(0,0)[r]{50000}}
\put(1116.0,594.0){\rule[-0.200pt]{4.818pt}{0.400pt}}
\put(220.0,690.0){\rule[-0.200pt]{4.818pt}{0.400pt}}
\put(198,690){\makebox(0,0)[r]{60000}}
\put(1116.0,690.0){\rule[-0.200pt]{4.818pt}{0.400pt}}
\put(220.0,786.0){\rule[-0.200pt]{4.818pt}{0.400pt}}
\put(198,786){\makebox(0,0)[r]{70000}}
\put(1116.0,786.0){\rule[-0.200pt]{4.818pt}{0.400pt}}
\put(220.0,113.0){\rule[-0.200pt]{0.400pt}{4.818pt}}
\put(220,68){\makebox(0,0){0.6}}
\put(220.0,766.0){\rule[-0.200pt]{0.400pt}{4.818pt}}
\put(335.0,113.0){\rule[-0.200pt]{0.400pt}{4.818pt}}
\put(335,68){\makebox(0,0){0.7}}
\put(335.0,766.0){\rule[-0.200pt]{0.400pt}{4.818pt}}
\put(449.0,113.0){\rule[-0.200pt]{0.400pt}{4.818pt}}
\put(449,68){\makebox(0,0){0.8}}
\put(449.0,766.0){\rule[-0.200pt]{0.400pt}{4.818pt}}
\put(564.0,113.0){\rule[-0.200pt]{0.400pt}{4.818pt}}
\put(564,68){\makebox(0,0){0.9}}
\put(564.0,766.0){\rule[-0.200pt]{0.400pt}{4.818pt}}
\put(678.0,113.0){\rule[-0.200pt]{0.400pt}{4.818pt}}
\put(678,68){\makebox(0,0){1}}
\put(678.0,766.0){\rule[-0.200pt]{0.400pt}{4.818pt}}
\put(792.0,113.0){\rule[-0.200pt]{0.400pt}{4.818pt}}
\put(792,68){\makebox(0,0){1.1}}
\put(792.0,766.0){\rule[-0.200pt]{0.400pt}{4.818pt}}
\put(907.0,113.0){\rule[-0.200pt]{0.400pt}{4.818pt}}
\put(907,68){\makebox(0,0){1.2}}
\put(907.0,766.0){\rule[-0.200pt]{0.400pt}{4.818pt}}
\put(1022.0,113.0){\rule[-0.200pt]{0.400pt}{4.818pt}}
\put(1022,68){\makebox(0,0){1.3}}
\put(1022.0,766.0){\rule[-0.200pt]{0.400pt}{4.818pt}}
\put(1136.0,113.0){\rule[-0.200pt]{0.400pt}{4.818pt}}
\put(1136,68){\makebox(0,0){1.4}}
\put(1136.0,766.0){\rule[-0.200pt]{0.400pt}{4.818pt}}
\put(220.0,113.0){\rule[-0.200pt]{220.664pt}{0.400pt}}
\put(1136.0,113.0){\rule[-0.200pt]{0.400pt}{162.126pt}}
\put(220.0,786.0){\rule[-0.200pt]{220.664pt}{0.400pt}}
\put(45,449){\makebox(0,0){$\sigma_c$}}
\put(678,23){\makebox(0,0){$M_H/M_W$}}
\put(325,450){\makebox(0,0)[l]{a}}
\put(335,546){\makebox(0,0)[l]{b,c}}
\put(220.0,113.0){\rule[-0.200pt]{0.400pt}{162.126pt}}
\multiput(220.59,769.31)(0.489,-1.951){15}{\rule{0.118pt}{1.611pt}}
\multiput(219.17,772.66)(9.000,-30.656){2}{\rule{0.400pt}{0.806pt}}
\multiput(229.58,736.23)(0.497,-1.625){51}{\rule{0.120pt}{1.389pt}}
\multiput(228.17,739.12)(27.000,-84.117){2}{\rule{0.400pt}{0.694pt}}
\multiput(256.58,650.24)(0.497,-1.318){49}{\rule{0.120pt}{1.146pt}}
\multiput(255.17,652.62)(26.000,-65.621){2}{\rule{0.400pt}{0.573pt}}
\multiput(282.58,583.51)(0.497,-0.930){51}{\rule{0.120pt}{0.841pt}}
\multiput(281.17,585.26)(27.000,-48.255){2}{\rule{0.400pt}{0.420pt}}
\multiput(309.58,534.09)(0.497,-0.752){49}{\rule{0.120pt}{0.700pt}}
\multiput(308.17,535.55)(26.000,-37.547){2}{\rule{0.400pt}{0.350pt}}
\multiput(335.58,495.27)(0.498,-0.699){103}{\rule{0.120pt}{0.658pt}}
\multiput(334.17,496.63)(53.000,-72.633){2}{\rule{0.400pt}{0.329pt}}
\multiput(388.58,421.83)(0.498,-0.528){103}{\rule{0.120pt}{0.523pt}}
\multiput(387.17,422.92)(53.000,-54.915){2}{\rule{0.400pt}{0.261pt}}
\multiput(441.00,366.92)(0.647,-0.498){79}{\rule{0.617pt}{0.120pt}}
\multiput(441.00,367.17)(51.719,-41.000){2}{\rule{0.309pt}{0.400pt}}
\multiput(494.00,325.92)(0.805,-0.497){63}{\rule{0.742pt}{0.120pt}}
\multiput(494.00,326.17)(51.459,-33.000){2}{\rule{0.371pt}{0.400pt}}
\multiput(547.00,292.92)(1.006,-0.497){49}{\rule{0.900pt}{0.120pt}}
\multiput(547.00,293.17)(50.132,-26.000){2}{\rule{0.450pt}{0.400pt}}
\multiput(599.00,266.92)(1.338,-0.496){37}{\rule{1.160pt}{0.119pt}}
\multiput(599.00,267.17)(50.592,-20.000){2}{\rule{0.580pt}{0.400pt}}
\multiput(652.00,246.92)(1.580,-0.495){31}{\rule{1.347pt}{0.119pt}}
\multiput(652.00,247.17)(50.204,-17.000){2}{\rule{0.674pt}{0.400pt}}
\multiput(705.00,229.92)(2.083,-0.493){23}{\rule{1.731pt}{0.119pt}}
\multiput(705.00,230.17)(49.408,-13.000){2}{\rule{0.865pt}{0.400pt}}
\multiput(758.00,216.92)(2.263,-0.492){21}{\rule{1.867pt}{0.119pt}}
\multiput(758.00,217.17)(49.126,-12.000){2}{\rule{0.933pt}{0.400pt}}
\multiput(811.00,204.92)(2.737,-0.491){17}{\rule{2.220pt}{0.118pt}}
\multiput(811.00,205.17)(48.392,-10.000){2}{\rule{1.110pt}{0.400pt}}
\multiput(864.00,194.93)(3.399,-0.488){13}{\rule{2.700pt}{0.117pt}}
\multiput(864.00,195.17)(46.396,-8.000){2}{\rule{1.350pt}{0.400pt}}
\multiput(916.00,186.93)(4.740,-0.482){9}{\rule{3.633pt}{0.116pt}}
\multiput(916.00,187.17)(45.459,-6.000){2}{\rule{1.817pt}{0.400pt}}
\multiput(969.00,180.93)(4.740,-0.482){9}{\rule{3.633pt}{0.116pt}}
\multiput(969.00,181.17)(45.459,-6.000){2}{\rule{1.817pt}{0.400pt}}
\multiput(1022.00,174.93)(5.831,-0.477){7}{\rule{4.340pt}{0.115pt}}
\multiput(1022.00,175.17)(43.992,-5.000){2}{\rule{2.170pt}{0.400pt}}
\multiput(1075.00,169.93)(5.831,-0.477){7}{\rule{4.340pt}{0.115pt}}
\multiput(1075.00,170.17)(43.992,-5.000){2}{\rule{2.170pt}{0.400pt}}
\put(1128.0,166.0){\rule[-0.200pt]{1.927pt}{0.400pt}}
\multiput(220.59,773.31)(0.489,-1.951){15}{\rule{0.118pt}{1.611pt}}
\multiput(219.17,776.66)(9.000,-30.656){2}{\rule{0.400pt}{0.806pt}}
\multiput(229.58,740.17)(0.497,-1.643){51}{\rule{0.120pt}{1.404pt}}
\multiput(228.17,743.09)(27.000,-85.087){2}{\rule{0.400pt}{0.702pt}}
\multiput(256.58,653.31)(0.497,-1.298){49}{\rule{0.120pt}{1.131pt}}
\multiput(255.17,655.65)(26.000,-64.653){2}{\rule{0.400pt}{0.565pt}}
\multiput(282.58,587.45)(0.497,-0.949){51}{\rule{0.120pt}{0.856pt}}
\multiput(281.17,589.22)(27.000,-49.224){2}{\rule{0.400pt}{0.428pt}}
\multiput(309.58,537.09)(0.497,-0.752){49}{\rule{0.120pt}{0.700pt}}
\multiput(308.17,538.55)(26.000,-37.547){2}{\rule{0.400pt}{0.350pt}}
\multiput(335.58,498.27)(0.498,-0.699){103}{\rule{0.120pt}{0.658pt}}
\multiput(334.17,499.63)(53.000,-72.633){2}{\rule{0.400pt}{0.329pt}}
\multiput(388.58,424.80)(0.498,-0.537){103}{\rule{0.120pt}{0.530pt}}
\multiput(387.17,425.90)(53.000,-55.900){2}{\rule{0.400pt}{0.265pt}}
\multiput(441.00,368.92)(0.647,-0.498){79}{\rule{0.617pt}{0.120pt}}
\multiput(441.00,369.17)(51.719,-41.000){2}{\rule{0.309pt}{0.400pt}}
\multiput(494.00,327.92)(0.805,-0.497){63}{\rule{0.742pt}{0.120pt}}
\multiput(494.00,328.17)(51.459,-33.000){2}{\rule{0.371pt}{0.400pt}}
\multiput(547.00,294.92)(1.006,-0.497){49}{\rule{0.900pt}{0.120pt}}
\multiput(547.00,295.17)(50.132,-26.000){2}{\rule{0.450pt}{0.400pt}}
\multiput(599.00,268.92)(1.338,-0.496){37}{\rule{1.160pt}{0.119pt}}
\multiput(599.00,269.17)(50.592,-20.000){2}{\rule{0.580pt}{0.400pt}}
\multiput(652.00,248.92)(1.580,-0.495){31}{\rule{1.347pt}{0.119pt}}
\multiput(652.00,249.17)(50.204,-17.000){2}{\rule{0.674pt}{0.400pt}}
\multiput(705.00,231.92)(1.929,-0.494){25}{\rule{1.614pt}{0.119pt}}
\multiput(705.00,232.17)(49.649,-14.000){2}{\rule{0.807pt}{0.400pt}}
\multiput(758.00,217.92)(2.477,-0.492){19}{\rule{2.027pt}{0.118pt}}
\multiput(758.00,218.17)(48.792,-11.000){2}{\rule{1.014pt}{0.400pt}}
\multiput(811.00,206.92)(2.737,-0.491){17}{\rule{2.220pt}{0.118pt}}
\multiput(811.00,207.17)(48.392,-10.000){2}{\rule{1.110pt}{0.400pt}}
\multiput(864.00,196.93)(3.399,-0.488){13}{\rule{2.700pt}{0.117pt}}
\multiput(864.00,197.17)(46.396,-8.000){2}{\rule{1.350pt}{0.400pt}}
\multiput(916.00,188.93)(3.465,-0.488){13}{\rule{2.750pt}{0.117pt}}
\multiput(916.00,189.17)(47.292,-8.000){2}{\rule{1.375pt}{0.400pt}}
\multiput(969.00,180.93)(4.740,-0.482){9}{\rule{3.633pt}{0.116pt}}
\multiput(969.00,181.17)(45.459,-6.000){2}{\rule{1.817pt}{0.400pt}}
\multiput(1022.00,174.93)(5.831,-0.477){7}{\rule{4.340pt}{0.115pt}}
\multiput(1022.00,175.17)(43.992,-5.000){2}{\rule{2.170pt}{0.400pt}}
\multiput(1075.00,169.94)(7.646,-0.468){5}{\rule{5.400pt}{0.113pt}}
\multiput(1075.00,170.17)(41.792,-4.000){2}{\rule{2.700pt}{0.400pt}}
\put(1128,165.67){\rule{1.927pt}{0.400pt}}
\multiput(1128.00,166.17)(4.000,-1.000){2}{\rule{0.964pt}{0.400pt}}
\multiput(220.58,758.77)(0.491,-1.798){17}{\rule{0.118pt}{1.500pt}}
\multiput(219.17,761.89)(10.000,-31.887){2}{\rule{0.400pt}{0.750pt}}
\multiput(230.58,724.03)(0.497,-1.688){49}{\rule{0.120pt}{1.438pt}}
\multiput(229.17,727.01)(26.000,-84.014){2}{\rule{0.400pt}{0.719pt}}
\multiput(256.58,638.24)(0.497,-1.318){49}{\rule{0.120pt}{1.146pt}}
\multiput(255.17,640.62)(26.000,-65.621){2}{\rule{0.400pt}{0.573pt}}
\multiput(282.58,571.51)(0.497,-0.930){51}{\rule{0.120pt}{0.841pt}}
\multiput(281.17,573.26)(27.000,-48.255){2}{\rule{0.400pt}{0.420pt}}
\multiput(309.58,522.16)(0.497,-0.732){49}{\rule{0.120pt}{0.685pt}}
\multiput(308.17,523.58)(26.000,-36.579){2}{\rule{0.400pt}{0.342pt}}
\multiput(335.58,484.30)(0.498,-0.689){103}{\rule{0.120pt}{0.651pt}}
\multiput(334.17,485.65)(53.000,-71.649){2}{\rule{0.400pt}{0.325pt}}
\multiput(388.58,411.83)(0.498,-0.528){103}{\rule{0.120pt}{0.523pt}}
\multiput(387.17,412.92)(53.000,-54.915){2}{\rule{0.400pt}{0.261pt}}
\multiput(441.00,356.92)(0.663,-0.498){77}{\rule{0.630pt}{0.120pt}}
\multiput(441.00,357.17)(51.692,-40.000){2}{\rule{0.315pt}{0.400pt}}
\multiput(494.00,316.92)(0.805,-0.497){63}{\rule{0.742pt}{0.120pt}}
\multiput(494.00,317.17)(51.459,-33.000){2}{\rule{0.371pt}{0.400pt}}
\multiput(547.00,283.92)(1.046,-0.497){47}{\rule{0.932pt}{0.120pt}}
\multiput(547.00,284.17)(50.066,-25.000){2}{\rule{0.466pt}{0.400pt}}
\multiput(599.00,258.92)(1.338,-0.496){37}{\rule{1.160pt}{0.119pt}}
\multiput(599.00,259.17)(50.592,-20.000){2}{\rule{0.580pt}{0.400pt}}
\multiput(652.00,238.92)(1.580,-0.495){31}{\rule{1.347pt}{0.119pt}}
\multiput(652.00,239.17)(50.204,-17.000){2}{\rule{0.674pt}{0.400pt}}
\multiput(705.00,221.92)(2.083,-0.493){23}{\rule{1.731pt}{0.119pt}}
\multiput(705.00,222.17)(49.408,-13.000){2}{\rule{0.865pt}{0.400pt}}
\multiput(758.00,208.92)(2.263,-0.492){21}{\rule{1.867pt}{0.119pt}}
\multiput(758.00,209.17)(49.126,-12.000){2}{\rule{0.933pt}{0.400pt}}
\multiput(811.00,196.93)(3.058,-0.489){15}{\rule{2.456pt}{0.118pt}}
\multiput(811.00,197.17)(47.903,-9.000){2}{\rule{1.228pt}{0.400pt}}
\multiput(864.00,187.93)(3.465,-0.488){13}{\rule{2.750pt}{0.117pt}}
\multiput(864.00,188.17)(47.292,-8.000){2}{\rule{1.375pt}{0.400pt}}
\multiput(917.00,179.93)(3.925,-0.485){11}{\rule{3.071pt}{0.117pt}}
\multiput(917.00,180.17)(45.625,-7.000){2}{\rule{1.536pt}{0.400pt}}
\multiput(969.00,172.93)(4.740,-0.482){9}{\rule{3.633pt}{0.116pt}}
\multiput(969.00,173.17)(45.459,-6.000){2}{\rule{1.817pt}{0.400pt}}
\multiput(1022.00,166.93)(5.831,-0.477){7}{\rule{4.340pt}{0.115pt}}
\multiput(1022.00,167.17)(43.992,-5.000){2}{\rule{2.170pt}{0.400pt}}
\multiput(1075.00,161.93)(5.831,-0.477){7}{\rule{4.340pt}{0.115pt}}
\multiput(1075.00,162.17)(43.992,-5.000){2}{\rule{2.170pt}{0.400pt}}
\put(1128.0,158.0){\rule[-0.200pt]{1.927pt}{0.400pt}}
\end{picture}

%% file: ipp.bbl
\begin{thebibliography}{9}
\bibitem{App75} T. Appelquist and J. Carrazzone, Phys. Rev. {\bf D11} (1975)
2856
\bibitem{Wei80} S. Weinberg, Phys. Lett. {\bf B91} (1980) 51
\bibitem{Geo93} H. Georgi, Ann. Rev. Nucl. Part. Sci. {\bf 43} (1993) 209
\bibitem{Bra95} E. Braaten, Phys. Rev. Lett. {\bf 74} (1995) 2164
\bibitem{Sin95} for recent developments, see "Electroweak Physics and the 
Early Universe", Eds. J.C. Romao and F. Freire, NATO ASI Series B:
Physics {\bf 338} (Plenum Press, 1995)
\bibitem{Jak96} A. Jakov\'ac, Phys. Rev. {\bf D53} (1996) 4538
\bibitem{Ker95} A. Kerres, G. Mack and G. Palma, Nucl. Phys. {\bf B467}
(1996) 510
\bibitem{Kaj95} K. Kajantie, M. Laine, K. Rummukainen and M. Shaposhnikov,
Nucl. Phys. {\bf B458} (1996) 90
\bibitem{Kaj96} K. Kajantie, M. Laine, K. Rummukainen and M. Shaposhnikov,
Nucl. Phys. {\bf B466} (1996) 189
\bibitem{Jak96b} A. Jakov\'ac, A. Patk\'os and P. Petreczky, Phys. Lett.
{\bf B367} (1996) 283
\bibitem{Arn93} P. Arnold and O. Espinosa, Phys. Rev. {\bf D47} (1993) 3546
\bibitem{Fod94} Z. Fodor and A. Hebecker, Nucl. Phys. {\bf B432} (1994) 127 
\bibitem{Bod94} D. B\"odeker, W. Buchm\"uller, Z. Fodor and T. Helbig, 
Nucl. Phys. {\bf B423} (1994) 171
\bibitem{Buc95} W. Buchm\"uller, Z. Fodor and A. Hebecker, Nucl. Phys. 
{\bf B447} (1995) 317
\bibitem{Far94} K. Farakos, K. Kajantie, K. Rummukainen and M. Shaposhnikov,
Nucl. Phys. {\bf B425} (1994) 67
\bibitem{Kaj93} K. Kajantie, K. Rummukainen and M. Shaposhnikov,
Nucl. Phys. {\bf B407} (1993) 356 
\bibitem{Los96} M. Losada, {\it High Temperature Dimensional Reduction of the
MSSM and other Multi-Scalar Models}, Rutgers-preprint RU-96-25
\bibitem{Cli96}J.M. Cline and K. Kainulainen, {\it Supersymmetric Electroweak 
Phase Transition Beyond Perturbation Theory} McGill-preprint, McGill/96-20
\bibitem{Lai96} M. Laine, {Effective Theories of MSSM at High Temperatures}
Heidelberg-preprint, HD-THEP-96-13
\bibitem{Fod95} Z. Fodor, J. Hein, K. Jansen, A. Jaster and I. Montvay, 
Nucl. Phys. {\bf B439} (1995) 147
\bibitem{Csi96} F. Csikor and Z. Fodor, private communication
\bibitem{KarTo} F. Karsch, T, Neuhaus, A. Patk\'os and J. Rank (in preparation)
 
\end{thebibliography}
